\documentclass[11pt]{article}

\usepackage{tablefootnote}
\usepackage[style=numeric,isbn=false,url=false,giveninits=true]{biblatex}
\usepackage{amsmath,hyperref, wasysym}
\usepackage{latexsym}
\usepackage{amssymb}
\usepackage[margin=0.75in]{geometry}
\usepackage{fancyhdr}
\usepackage{graphicx}
\usepackage{times}
\usepackage{color,soul}
\usepackage{cancel}
\usepackage{multicol}
\usepackage{placeins}
\usepackage{authblk}
\usepackage{arydshln}
\usepackage{float}
\usepackage{dsfont}
\usepackage{booktabs}
\usepackage{enumerate}
\usepackage{enumitem}
\usepackage{adjustbox}
\restylefloat{table}
\usepackage{lipsum}

\newcommand{\btheta}{\boldsymbol{\theta}}

\begin{document}
\title{Multi-objective Bayesian inference in an agent-based model of zebrafish patterns via topological data analysis}
\author[1]{Yue Liu}
\author[1]{Alexandria Volkening}

\affil[1]{Department of Mathematics, Purdue University, West Lafayette, IN. Email addresses for correspondence: liu4194@purdue.edu and avolkening@purdue.edu}


\maketitle

\begin{abstract}Spatial patterns arising from the collective behavior of individual agents are present across biological systems. While agent-based models offer a natural framework for uncovering unknown agent (e.g., cell) interactions, these stochastic models face significant challenges. For spatial patterns, agent-based modeling often involves manual tuning to attain qualitative consistency with multiple experiments. This process limits predictive power and raises questions about parameter identifiability and model uniqueness. Combining topological techniques and Bayesian computation, we present a multi-objective methodology for parameter inference in detailed models. We illustrate our approach by inferring parameters in an agent-based model of zebrafish patterns, achieving practical identifiability in several case studies. By introducing extended prior distributions, we then reframe parameter inference as \textit{rule inference}, allowing us to search across over 80 candidate agent-based rules to identify an alternative, simpler model consistent with our data.
\end{abstract}


\section{Introduction}

Spatial self-organization is widespread in nature, with examples including cell behavior during development \cite{buttenschonBridgingSingleCollective2020,patterson2019ZebrafishPigmentPattern} and animal flocking or schooling dynamics \cite{Mogilner1999,Lukeman2010keshet,Katz2011fish}. Working alongside empirical data, mechanistic modelers strive to tease out the behaviors driving pattern formation. Agent-based models, which treat individuals (e.g., cells) as agents interacting through rules and equations, are a natural framework for doing so and can provide deep insight. However, as models become more detailed, complexity grows. Researchers may seek to balance this complexity and increase predictive power by requiring model consistency with multiple experiments and more data \cite{SegelEdelsteinKeshet}. Often this balancing process is simplified and done by hand, in part because it is difficult to quantify messy spatial patterns, and in part because agent-interaction rules---not just parameter values---may be unknown. At the same time, rich data (on variability, patterning timelines, and multiple experiments) are increasingly available \cite{Baker2025Rev}, and we want to build models that are consistent with these data, and determine if alternative models or parameters would also work. This is intractable by hand, due to constraints on the number of patterns, parameters, and rules a modeler can evaluate manually, leading to long-standing questions about model uniqueness and parameter identifiability. With this motivation, here we propose multi-objective Bayesian inference, driven by topological data analysis, as a tractable means of performing data-driven, biologically detailed agent-based modeling. To ground our approach in practical challenges, we develop and illustrate it using an agent-based model \cite{volkeningIridophoresSourceRobustness2018} of zebrafish skin patterns.

\begin{figure*}[t!]
    \centering
    \includegraphics[width=0.9\textwidth]{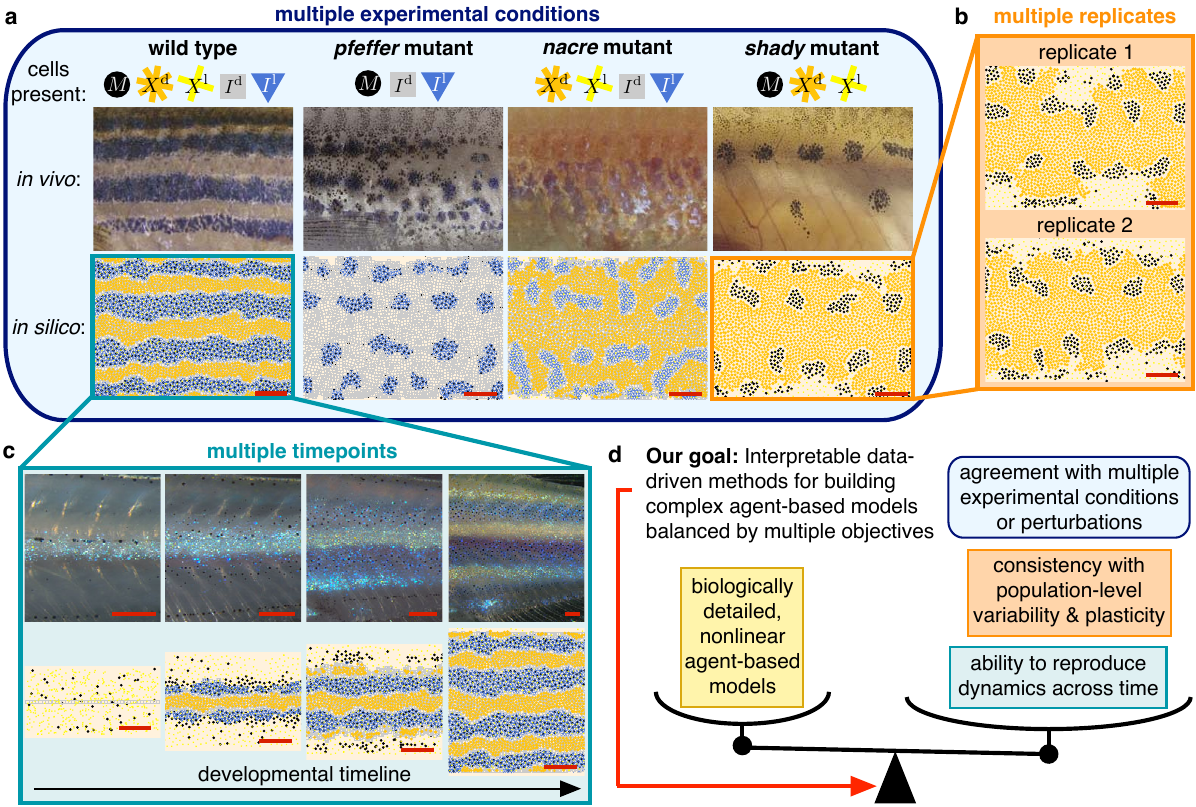}
    \caption{Overview of our motivation: balancing biologically sophisticated agent-based models with multiple objectives. 
    (a) Building complex models of pattern formation often involves seeking to replicate multiple experimental conditions \cite{SegelEdelsteinKeshet}. In the focal example of our study, these conditions are wild-type zebrafish stripes and mutant patterns that emerge because specific cell types are missing (but the remaining cells seem to act normally) \cite{frohnhoferIridophoresTheirInteractions2013}. (b) The biologically detailed ABM \cite{volkeningIridophoresSourceRobustness2018} describes the stochastic cell interactions behind patterning. Together with newly emerging data, stochastic ABMs offer the opportunity to capture not just large-scale, characteristic features, but also variability across replicates and messiness within individual patterns \cite{khoudari2025,mcguirlTopologicalDataAnalysis2020} (c) While the images in (a, b) present juvenile or adult fish, patterning is a dynamic process \cite{frohnhoferIridophoresTheirInteractions2013}. (d) Across biological systems, ABMs are simultaneously known for their intractability---whether in terms of analysis, inference, or quantification---and for their ability to provide valuable insight when well calibrated. However, agent-based modeling has traditionally depended on manual tuning and qualitative inspection of patterns, limiting perspective and raising questions about practical identifiability. We help address this challenge by combining Bayesian inference methods \cite{buzbasAABCApproximateApproximate2015} with topological data analysis \cite{Munch2020Shape,Otter2017mason}. Our data-driven approach quantitatively expands on modeler intuition, drawing together multiple sources of data to constrain unknowns in detailed models. Empirical images in (a, c) are adapted from Frohnh\"{o}fer \textit{et al.} \cite{frohnhoferIridophoresTheirInteractions2013}, published by The Company of Biologists Ltd., and licensed under CC-BY 3.0 (\protect \url{http://creativecommons.org/licenses/by/3.0}).}
    \label{fig:zebrafish_picture}
\end{figure*}

Zebrafish \textit{(Danio rerio)} is a model organism for studying pattern formation \cite{kondo2021StudiesTuringPattern,patterson2019ZebrafishPigmentPattern} (Fig.~\ref{fig:zebrafish_picture}). Over several weeks, its black and gold stripes emerge in the growing skin due to the migration, differentiation, division, and competition of pigment cells, relying on local and long-range cell interactions \cite{Jan,Yamaguchi}. While wild-type patterns are made up of three main cell types, three mutants have spotty patterns that each lack one of the main cell populations. Experiments suggest that the patterns in these mutants---\textit{pfeffer} \cite{Maderspacher2003,ParTur130,PatDev127}, \textit{nacre} \cite{Lister,Maderspacher2003}, and \textit{shady} \cite{Lopes,frohnhoferIridophoresTheirInteractions2013}---are altered simply because they are missing cells. In contrast, different patterns arise in many other mutants due to changes in cell behavior. To shed light on poorly understood cell interactions, Volkening and Sandstede \cite{volkeningIridophoresSourceRobustness2018} built an agent-based model (ABM). Their stochastic, off-lattice model reproduces the developmental timeline of pattern formation on growing domains, and makes testable predictions about 
cell interactions (Fig.~\ref{fig:abm_schematic}). Because the broad types of agent behavior in the ABM \cite{volkeningIridophoresSourceRobustness2018}---movement, birth, death, and type changes---are present in other systems as well, using this model to motivate our inference methodology provides a concrete example
that suggests wider applicability.

For biologically detailed models, incorporating multiple experiments into the process of 
model building, parameter estimation, and model validation 
is natural \cite{SegelEdelsteinKeshet}. Depending on the system, ``multiple experiments" take different forms. For our focal ABM \cite{volkeningIridophoresSourceRobustness2018}, wild-type, \textit{pfeffer}, \textit{nacre}, and \textit{shady} patterns serve as multiple experiments (Fig.~\ref{fig:zebrafish_picture}(a)). The authors \cite{volkeningIridophoresSourceRobustness2018} constrained unknown rules of cell behavior by requiring qualitative consistency with these four conditions, where the only change made from wild type to simulate each mutant was to turn off birth for the appropriate cell type. In other settings, such as wound-healing assays, multiple experiments can mean different initial conditions (i.e., scratch profiles) \cite{Perez2022}. The key feature in either case is that the relationship between the experiments is known, and the goal is to find one model that is consistent across all of them. Because adjusting a nonlinear model to fit one experiment affects agreement with other data, reproducing multiple experiments---e.g., meeting multiple objectives---is a strong constraint \cite{SegelEdelsteinKeshet}.

Despite the value of calibrating models across multiple experiments, specifying parameter values in ABMs is often a manual, qualitative process. This stems from two challenges: first, while inference methods are well established for population-density models \cite{Simpson2025Rev}, less research has considered parameter inference for ABMs, especially for detailed models. Because calculating a likelihood function for complex ABMs is typically intractable, likelihood-free approaches, such as approximate Bayesian computation (ABC) \cite{sunnaker2013ApproximateBayesianComputation,Beaumont2019Rev} and related methods \cite{buzbasAABCApproximateApproximate2015,Lambert2018,Jorgensen2022} are necessary. Second, inference techniques
rely on having a means to quantitatively judge model and data fit, but this is not straightforward for spatial, point-cloud patterns. The community has long addressed this by simplifying to non-spatial summary statistics, such as the number of cells in time. Moreover, in many cases, parameter inference for ABMs has considered single experiments, and we highlight refs.\ \cite{Bergman2024, Perez2022,ciocanel2025enhancinggeneralizabilitymodeldiscovery} as recent studies that shift toward a more multi-objective perspective. While Bergman et al. \cite{Bergman2024}
and Ciocanel et al. \cite{ciocanel2025enhancinggeneralizabilitymodeldiscovery} inferred parameters in on-lattice ABMs based on non-spatial summary statistics, Martina Perez et al.\ \cite{Perez2022} used pair correlation functions, density profiles, and cell counts to estimate parameters across variable initial conditions for an ABM of one cell population in wound-healing assays. 
These studies \cite{Bergman2024,Ciocanel2021,Perez2022} point to the value of a multi-experiment approach to inference for ABMs.
However, many of the models or patterns considered previously are less complex than our focal ABM \cite{volkeningIridophoresSourceRobustness2018}, and non-spatial analysis is not appropriate for zebrafish patterns.

Over the last decade, topological data analysis (TDA) has emerged as a powerful means of providing insight into spatial patterns. Persistent homology can describe the ``shape" of point-cloud data by tracking topological features (connected components, loops, and higher-dimensional analogues) across scales from single point to collective group \cite{Munch2020Shape,Edelsbrunner2008,Otter2017mason}. Well suited for agent-based data, persistent homology has been applied widely \cite{bhaskar2019AnalyzingCollectiveMotion,clevelandQuantifyingDifferentModeling2023, mcguirlTopologicalDataAnalysis2020}, and it has recently been combined with ABC and related methods under single objectives \cite{thorneTopologicalApproximateBayesian2022,wenzel2025TopologicallybasedParameterInference,McDonald2025,jin2025topology}. Several studies \cite{jin2025topology,thorneTopologicalApproximateBayesian2022,McDonald2025} have inferred parameters based on the shape of data at the final simulation time, whereas others \cite{wenzel2025TopologicallybasedParameterInference} have considered TDA in time for deterministic ABMs that do not include changes in population size. Together refs.\ \cite{thorneTopologicalApproximateBayesian2022,wenzel2025TopologicallybasedParameterInference,McDonald2025,jin2025topology} demonstrate the potential for joint TDA--ABC methods to address long-standing challenges in parameter inference for ABMs. They also raise questions about whether or not it is feasible to develop TDA-based inference approaches for more detailed models that seek to replicate multiple experiments and time dynamics.

Beyond the challenges associated with parameter inference and quantification for ABMs, another reason that mechanistic modelers often work by hand is that sometimes not just the parameters, but also the model rules, are unknown. For example, it can be unclear what types of cells are communicating with what other types of cells at what distance away. 
Then the bottleneck question is not ``what are the parameter values?”, but ``what are the interaction rules?”. For population-level and differential-equation models, these questions are intertwined in the fields of model selection \cite{liu2024parameter,Baker2025Rev,wenzel2025TopologicallybasedParameterInference,McDonald2025,mangan2017model} and equation learning \cite{nardini2021learning,ciocanel2025enhancinggeneralizabilitymodeldiscovery,brunton2016discovering,raissi_hidden_2018,lagergren_biologically-informed_2020,Kevrekids}. Considering $\frac{dx}{dt} = f(x; \theta)$, one goal is discovering the form of $f(\cdot)$ by selecting from a library of possible terms with sparsity constraints. Could something conceptually similar apply to ABMs with rules for agent interactions, rather than differential equations? Being able to infer the rules governing cell behavior---as well as the parameters in those rules---would substantially increase the predictive power of mechanistic ABMs.

Inspired by the effectiveness of topological techniques for parameter inference \cite{wenzel2025TopologicallybasedParameterInference,thorneTopologicalApproximateBayesian2022,McDonald2025}, here we unite approximate Bayesian inference methods, TDA, and a multi-objective perspective to corral and calibrate detailed ABMs of pattern formation. 
In a quantitative, interpretable, computationally feasible way, our methodology mirrors the mechanistic modeler's approach of carefully setting parameters and rules to achieve consistency across experiments and time \cite{SegelEdelsteinKeshet}. Motivating our work using an ABM \cite{volkeningIridophoresSourceRobustness2018} for zebrafish patterns, we find that single objectives---e.g., one phenotype at the final simulation time---are insufficient to achieve practical identifiability. Instead, combining just a few objectives allows us to accurately infer parameters. Moreover, as a step toward data-driven inference of ABM rules---not just parameters---we conclude by showing how our methodology can identify redundant cell interactions and suggest alternative, simpler models. Taken together, our main contribution is a biologically grounded methodology for multi-objective inference in stochastic ABMs of complex spatial patterning. Our work contributes to long-standing questions about uniqueness for computational models, and, as a step toward ``interaction rule learning", it opens up opportunities for data-driven discovery of cell behaviors.

\section{Results: Our multi-objective, TDA-based approach}\label{sec:resulst1}

\begin{figure}
    \centering
    \includegraphics[width=0.5\textwidth]{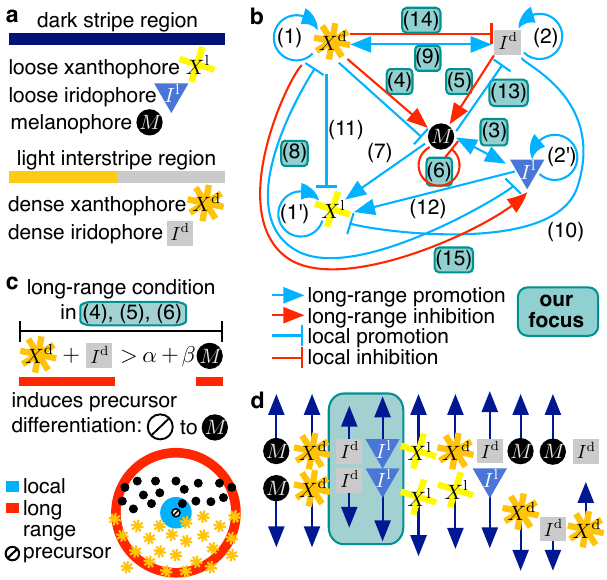}
    \caption{Overview of our focal ABM \cite{volkeningIridophoresSourceRobustness2018}. (a) Dark stripes and spots in zebrafish skin consist of $X^\text{l}$, $I^\text{l}$, and $M$ cells, while light regions contain $X^\text{d}$ and $I^\text{d}$ cells. (b) The ABM \cite{volkeningIridophoresSourceRobustness2018} was developed to address current unknowns about iridophores, and other cell interactions were based in the biological literature (e.g., refs.\ \cite{Singh,PatPLos,Mcmen2014,Mahalwar,nakamasu2009InteractionsZebrafishPigment,walderich2016homotypic,PatDev127}). This schematic summarizes signals involved in differentiation, division, competition, and transitions in type. (c) Except for migration, cell dynamics are modeled as stochastic, discrete-time rules involving local and long-range neighborhoods \cite{volkeningIridophoresSourceRobustness2018}. To develop our inference methodology, we first consider long-range signals for $M$ differentiation. At each simulated day of development, precursor positions are selected uniformly at random and evaluated for possible $M$ differentiation \cite{volkeningIridophoresSourceRobustness2018}. If the number of $X^\text{d}$ and $I^\text{d}$ cells in an annulus around a precursor is high, the precursor becomes an $M$ cell. We infer the time delay $\alpha$ and scaling parameter $\beta$. (d) Cell migration is governed by coupled ordinary differential equations (ODEs), based largely on repulsive forces. As our second test case, we infer two parameters involved in iridophore migration, before turning to a study of six parameters in poorly understood cell interactions (Fig.~\ref{fig:irid-params-vary-acceptance}).
    }
    \label{fig:abm_schematic}
\end{figure}

Here we apply TDA  \cite{Munch2020Shape,Otter2017mason,mcguirlTopologicalDataAnalysis2020} to quantify pattern variability, and then introduce our pipeline for parameter inference and rule selection in complex ABMs. To ground our work in concrete challenges, we develop it using the stochastic, off-lattice ABM  \cite{volkeningIridophoresSourceRobustness2018}. This is an established, detailed model, depending on over sixty parameters to describe the interactions of thousands of cells of five types, which we refer to as $M$, $X^\text{d}$, $X^\text{l}$, $I^\text{d}$, and $I^\text{l}$ cells; see Fig.~\ref{fig:abm_schematic}. While we frame our approach around zebrafish, we expect it to be applicable to many other settings with agent movement, birth, death, and transitions in type.

\subsection{Quantifying the distance between pattern replicates through TDA}

\begin{figure*}[t!]
    \centering
    \includegraphics[width=\textwidth]{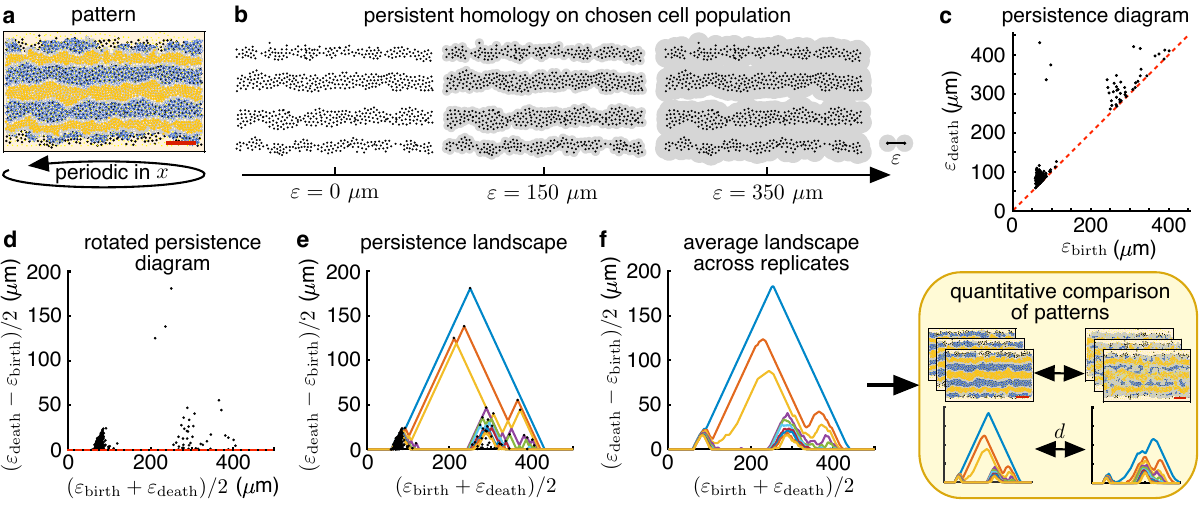} \vspace{-\baselineskip}
    \caption{Overview of TDA in our inference methodology. (a) We enforce periodic boundary conditions horizontally \cite{volkeningIridophoresSourceRobustness2018}, and (b) apply persistent homology to cell positions under the Vietoris--Rips filtration, considering each cell type independently. Computing persistent homology can be thought of as allowing a ball with diameter $\varepsilon$ centered at each cell to grow. Balls slowly overlap with one another, and we track the connected components (dimension-$0$ features) and loops (dimension-$1$ features) as $\varepsilon$ grows. Due to periodicity, stripes lead to loops that persist for a wide $\varepsilon$ range, while spots are persistent connected components \cite{mcguirlTopologicalDataAnalysis2020,clevelandQuantifyingDifferentModeling2023}. (c) Persistence diagrams summarize the shape of data in a given dimension (here dimension $1$): each point $(\varepsilon_\text{birth},\varepsilon_\text{death}$) denotes the value $\varepsilon$ at which a feature emerges and disappears. (d) To make the output of persistent homology amenable for averaging across replicates, we rotate persistence diagrams and (e) construct a family of curves called a persistence landscape \cite{bubenik2015StatisticalTopologicalData,bubenik2017PersistenceLandscapesToolbox}. (f) Persistence landscapes can be discretized and averaged, allowing us to compare ground-truth patterns to stochastic simulations \cite{volkeningIridophoresSourceRobustness2018} by measuring the distance between their mean persistence landscapes.}
    \label{fig:tda_schematic}
\end{figure*}

As an important ingredient in our inference pipeline, we seek to determine whether models are consistent with the timeline of zebrafish patterning (Fig.~\ref{fig:whole_pipeline}(c)) and with variable, messy features across replicates (Fig.~\ref{fig:whole_pipeline}(b)). Prior work \cite{mcguirlTopologicalDataAnalysis2020,clevelandQuantifyingDifferentModeling2023} has demonstrated the utility of TDA, specifically persistent homology, in quantifying \textit{in silico} zebrafish patterns at the juvenile stage, and we build on this to characterize patterns across developmental time. 

Given a pattern, we compute persistent homology by selecting a focal cell type (e.g., $M$ in Fig.~\ref{fig:tda_schematic}) and constructing simplicial complexes using the Vietoris--Rips filtration. Informally, this involves centering a ball of diameter $\varepsilon$ on the coordinates of each $M$ cell, connecting two cells within a distance $\varepsilon$ apart, and filling in the triangle between three cells when their $\varepsilon$-balls pairwise intersect. (See refs.\ \cite{topaz2015TopologicalDataAnalysis,Edelsbrunner2008,Otter2017mason} for details.) As $\varepsilon$ grows, we focus on the evolution of $0$- and $1$-dimensional topological features. The scale $\varepsilon = \varepsilon_\text{birth}$ at which a feature appears is its ``birth time", the scale $\varepsilon = \varepsilon_\text{death}$ at which it disappears is its ``death time", and $\varepsilon_\text{death}-\varepsilon_\text{birth}$ is its ``persistence" \cite{topaz2015TopologicalDataAnalysis}. Because we enforce periodic boundaries horizontally, stripes are associated with persistent dimension-$1$ features, while spots are reflected in connected components \cite{mcguirlTopologicalDataAnalysis2020,clevelandQuantifyingDifferentModeling2023}.

The results of persistent homology can be summarized in a persistence diagram (Fig.~\ref{fig:tda_schematic}(c)), where each point $(\varepsilon_\text{birth},~\varepsilon_\text{death})$ denotes a topological feature. Because these diagrams are not amenable to statistical analysis, we transform them to persistence landscapes (Fig.~\ref{fig:tda_schematic}(d)--(e)). The value of persistence landscapes in the case of biological data is well established \cite{Hartsock2025,Stolz2021PL}. Introduced by Bubenik \cite{bubenik2015StatisticalTopologicalData}, a persistence landscape is a vectorized representation of a persistence diagram that can be averaged uniquely across replicates. This quality is key: because biological patterns are messy and the ABM \cite{volkeningIridophoresSourceRobustness2018}  is stochastic, we perform repeated simulations under each parameter proposal $\btheta^*$ before judging its fit with data. We use $N_\text{rep} =5$ replicates, and the mean persistence landscape across those five replicates serves as the quantitative signature of the model under $\btheta^*$ (at one time). We can then measure the distance between two sets of pattern replicates as the Euclidean distance $d$ between their associated mean landscapes; see Methods. As we show in Fig.~\ref{fig:m-birth-tda-vs-pattern}, patterns with low $d$ are qualitatively similar to our ground-truth data.

\subsection{Our TDA-based methodology for multi-objective inference}

\begin{figure}[t!]
    \centering
    \includegraphics[width=0.5\textwidth]{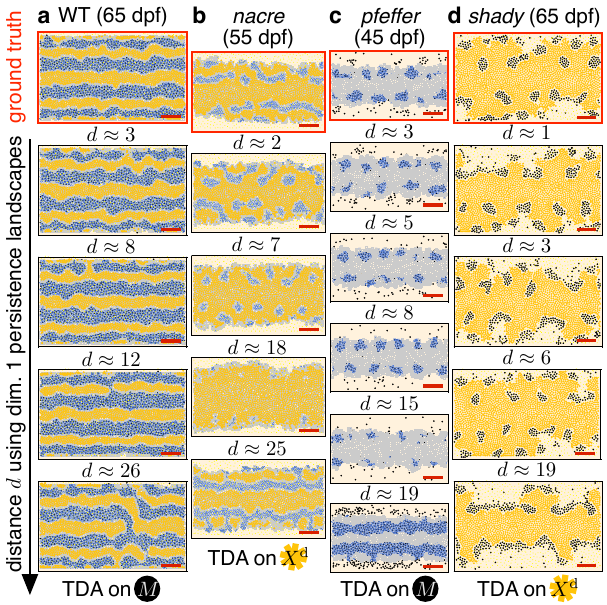}
    \caption{Measuring the distance between pattern replicates using a TDA-based approach. Each distance $d$ is based on simulating the ABM \cite{volkeningIridophoresSourceRobustness2018} $N_\text{rep}=5$ times under a given set of parameter values, computing persistent homology for the indicated cell type at the selected time, constructing persistence landscapes, and discretizing the mean landscapes to create vector summaries \cite{bubenik2017PersistenceLandscapesToolbox}; $d$ is the Euclidean distance between the vector summaries for our ground-truth data and simulations. As in refs.\ \cite{mcguirlTopologicalDataAnalysis2020,clevelandQuantifyingDifferentModeling2023}, we remove cells in the top and bottom $10$\% of the domain before applying TDA to focus on the portion of the pattern that is most well formed. We show (a) wild-type, (b) \textit{nacre}, (c) \textit{pfeffer}, and (d) \textit{shady} patterns under the baseline parameter values in ref.~\cite{volkeningIridophoresSourceRobustness2018}, followed by patterns under altered parameters that give rise to a range of $d$ values (see Supporting Information~\ref{apx:fig_detail}).
    High distances coincide with poor qualitative fits.}
    \label{fig:m-birth-tda-vs-pattern}
\end{figure}

\begin{figure*}[t!]
    \centering
    \includegraphics[width=\textwidth]{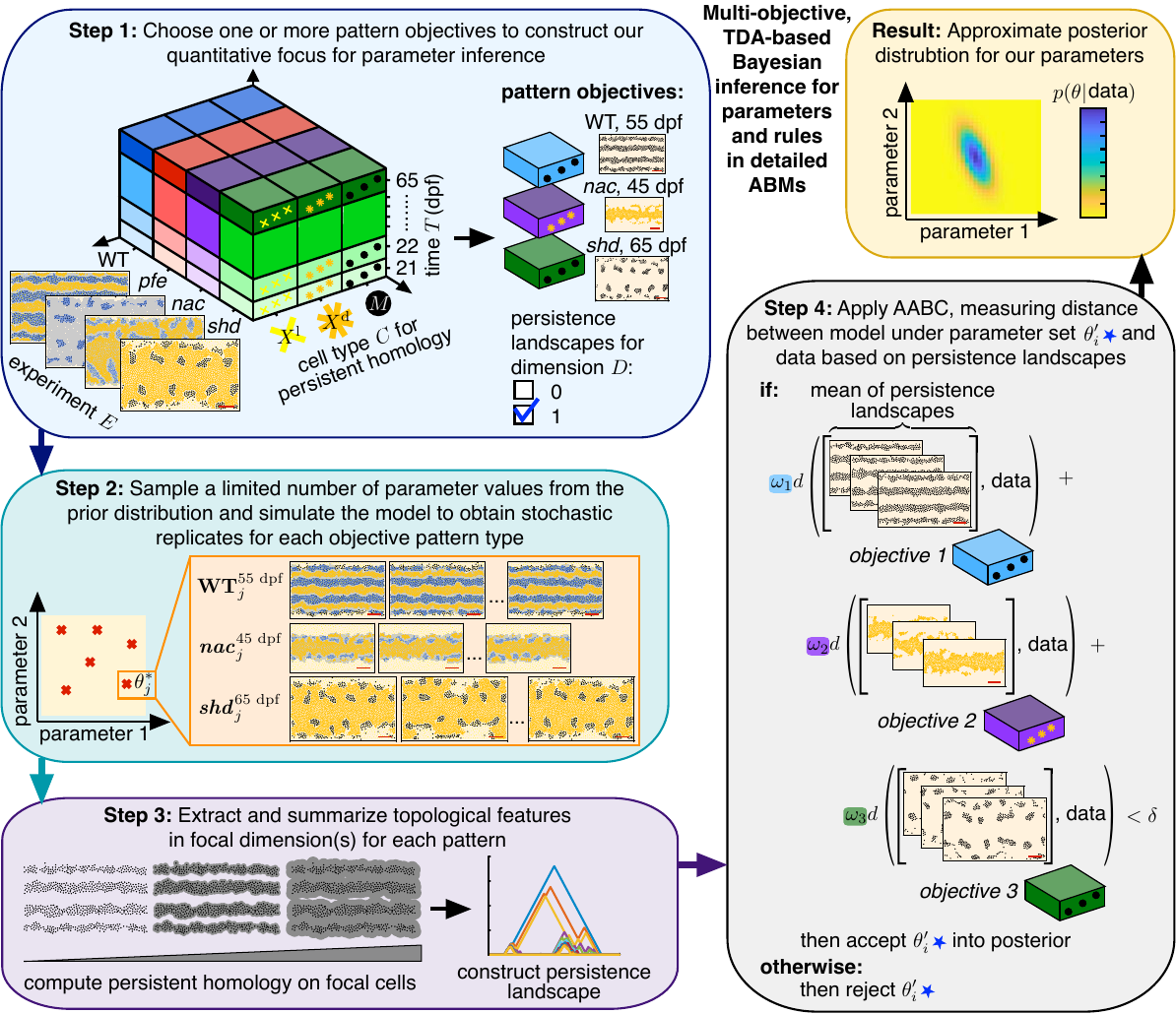} \vspace{-\baselineskip}
    \caption{Our multi-objective inference methodology. Step~$1$ involves choosing the perspective(s) that we will take on our biological system. In the case of zebrafish patterns, this means choosing between four experimental conditions, selecting developmental time point(s) to analyze, and determining what cell population(s) and topological dimension(s) to focus on when quantifying patterns using persistence landscapes. Step~$2$ implements the pre-proposal stage of AABC; see Methods and Fig.~\ref{fig:aabc_schematic}. We sample possible parameter values from the prior distribution and simulate the stochastic model \cite{volkeningIridophoresSourceRobustness2018} repeatedly to produce multiple replicates of each pattern corresponding to our selected objectives. In Step~$3$, we apply TDA, replacing patterns with persistence landscapes; see Fig.~\ref{fig:tda_schematic}. Step~$4$ implements the pseudo-proposal stage of AABC. For each objective (i.e., choice of experiment $E \in \{\text{wild type}, \textit{pfeffer}, \textit{nacre}, \textit{shady}\}$, time point $T \in \{21, 22, \dots, 65\}$ dpf, cell type $C \in \{M,X^\text{l},X^\text{d}\}$, and topological dimension $D\in \{0,1\}$), we measure the difference between pattern replicates and data as the distance between their mean persistence landscapes. To account for multiple objectives, we take a weighted sum of these distances and accept a parameter proposal if its model--data difference is low. This leads to an estimate of the posterior distribution for our parameters.}
    \label{fig:whole_pipeline}
\end{figure*}

Our approach for parameter inference and ABM rule selection draws together approximate approximate Bayesian computation (AABC) \cite{buzbasAABCApproximateApproximate2015,Lambert2018} and TDA in a multi-objective way through four steps (Fig.~\ref{fig:whole_pipeline}).
We adopt a Bayesian viewpoint, so we seek the posterior distribution $\text{Pr}(\btheta | \textbf{x}_\text{data})$ of our parameters~$\btheta$ given data $\textbf{x}_\text{data}$. This is related to the likelihood $\text{Pr}(\textbf{x}_\text{data} |\btheta )$ and prior distribution $\text{Pr}(\btheta)$ as $\text{Pr}(\btheta | \textbf{x}_{\textrm{data}}) \propto \text{Pr}(\textbf{x}_{\textrm{data}} |\btheta) ~\text{Pr}(\btheta)$ by Bayes' rule. Here $\btheta$ is multidimensional, and simulations of the model \cite{volkeningIridophoresSourceRobustness2018} under its baseline parameters serve as ground-truth data. Using \textit{in silico} data allows us to test the accuracy of our methods in this first study.

In Step~$1$, we choose objectives, selecting how we will judge the model. To do so it is useful to list all of the perspectives that we could take, considering experiments, times, and cell types. First, the ABM \cite{volkeningIridophoresSourceRobustness2018} was developed to replicate four phenotypes: wild-type stripes and messy spots in 
three mutants (\textit{pfeffer}, which lacks $M$ cells; \textit{nacre}, which lacks $X^\text{d}$ and $X^\text{l}$; and \textit{shady}, which lacks $I^\text{d}$ and $I^\text{l}$ \cite{Maderspacher2003,ParTur130,PatDev127,Lister,frohnhoferIridophoresTheirInteractions2013}). These phenotypes serve as four experimental conditions, differing only by turning off differentiation for the missing cells \cite{volkeningIridophoresSourceRobustness2018}. Next, for each condition, the ABM \cite{volkeningIridophoresSourceRobustness2018} describes patterning from about $21$ to $65$~days post fertilization (dpf). And lastly, at each time, we must select a cell type and dimension to focus on when computing persistent homology. Each combination of experiment $E \in \{\text{wild type}, \textit{pfeffer}, \textit{nacre}, \textit{shady}\}$, time $T \in \{21, \dots,65\}$ dpf, cell type $C \in \{M, X^\text{d}, X^\text{l}\}$, and topological dimension $D \in \{0, 1\}$ represents a single perspective on zebrafish that gives rise to one objective. (As a simplification, we do not apply TDA to $I^\text{d}$ or $I^\text{l}$.) Step~$1$ involves selecting $N_\text{obj} \ge 1$ such objectives for assessing model--data fit.

Step~$2$ implements the pre-proposal stage of AABC \cite{buzbasAABCApproximateApproximate2015,Lambert2018}; see Methods and Fig.~\ref{fig:aabc_schematic}. This means sampling a limited number $N_\text{pre}$ of parameter values $\{\btheta^*\}$ from our prior distribution $\text{Pr}(\btheta)$. We then repeatedly simulate the ABM \cite{volkeningIridophoresSourceRobustness2018} under each parameter proposal $\btheta^*$ to produce $N_\text{rep} = 5$~pattern timelines per experiment $E$ that we select in Step~$1$, accounting for stochasticity. Importantly, we specify broad, uniform priors. This illustrates the tractability of our approach when little is known about the parameters, and it is part of what allows us to shift from inferring parameters to selecting model rules.

TDA enters our methodology in Step~$3$ (Fig.~\ref{fig:whole_pipeline}) to quantify patterns across scales, from cell to tissue. For each parameter proposal $\btheta^*$, viewed from a single perspective $(E, T, C, D)$, we associate it with $N_\text{rep}$ persistence landscapes, each describing one of its five stochastic replicates. Our quantitative summary of the patterns generated by the ABM \cite{volkeningIridophoresSourceRobustness2018} under the parameter $\btheta^*$ thus takes the form of $N_\text{rep} \times N_\text{obj}$ persistence landscapes. Together Steps $1$--$2$ are the most computationally costly part of our  methodology. To help others gauge the feasibility of applying our pipeline to their systems, simulating the ABM \cite{volkeningIridophoresSourceRobustness2018} once, computing persistent homology, and building persistence landscapes for each $C \in \{M, X^\text{d}, X^\text{l}\}
$, $T \in \{21,\dots,65\}$, and $D \in \{0, 1\}$ takes roughly $5$ minutes on a standard desktop computer.

In Step~$4$, we build on the AABC approach of Buzbas and Rosenberg \cite{buzbasAABCApproximateApproximate2015} to account for multiple objectives. The key idea in AABC is augmenting costly simulations of one's focal model with pseudo-simulations from a statistical surrogate \cite{buzbasAABCApproximateApproximate2015,Lambert2018}. Here we again sample from the prior, now selecting many ($N_\text{pseu} \gg N_\text{pre}$) values. Our goal is to assign each pseudo-proposal $\btheta'$ a set of $N_\text{obj}$ mean persistence landscapes, without simulating the ABM \cite{volkeningIridophoresSourceRobustness2018} or computing persistent homology. Instead, we find the nearest $N_\text{rep}$ proposals $\{\btheta^*_i\}$ to $\btheta'$ \cite{buzbasAABCApproximateApproximate2015}, noting that each proposal $\btheta^*_i$ is already associated with $N_\text{rep}$ landscapes per objective from Step~$3$. Weighting by distance in parameter space as in ref.\ \cite{buzbasAABCApproximateApproximate2015}, we sample from these $N_\text{rep}^2$ landscapes per objective to assign $\btheta'$ a set of $N_\text{rep} \times N_\text{obj}$ persistence landscapes (Fig.~\ref{fig:aabc_schematic} and Methods). We then compute the mean landscape per objective 
to estimate what the quantitative output of the ABM would be under $\btheta'$.

To complete Step~$4$, we determine whether to accept the parameter value $\btheta'$ into the posterior, using the distance between its mean persistence landscapes and the mean landscapes for our ground-truth data. As in Fig.~\ref{fig:whole_pipeline}, we take a weighted sum of these distances across our objectives: if the weighted sum is small, we accept $\btheta'$ into the posterior. Repeating this process many times leads to an estimate of the posterior distribution of the parameters. Notably, we could choose to weight good performance on wild type more than good performance on mutants, or to prioritize the final simulation time. Adding complexity, different experiments and times admit a range of distance values (Fig.~\ref{fig:m-birth-tda-vs-pattern}). For each perspective $(E,T, C,D)$, we thus set its weight $w$ to be the inverse of the minimum distance $d(\btheta^*;E, T, C, D)$ between model output and data, evaluated across the pre-proposals $\{\btheta^*\}$ from Steps~$1$--$2$. This means that each term in the weighted sum captures error in multiples of the minimum error observed. 

\section{Results: An inference case study of zebrafish patterns}\label{sec:results}

We now demonstrate our parameter inference and rule selection pipeline through three case studies of the ABM~\cite{volkeningIridophoresSourceRobustness2018}. Our first two examples---focused on a stochastic rule for cell proliferation and differential equations for cell migration---highlight the importance of a multi-objective perspective when calibrating complex models. We conclude with a study of six parameters in the ABM rules for cell-type transitions. Adjusting these parameters to extreme values turns off different signaling mechanisms, and our methods are able to select simplified model rules.

\begin{figure*}[t!]
    \centering
    \includegraphics[width=\textwidth]{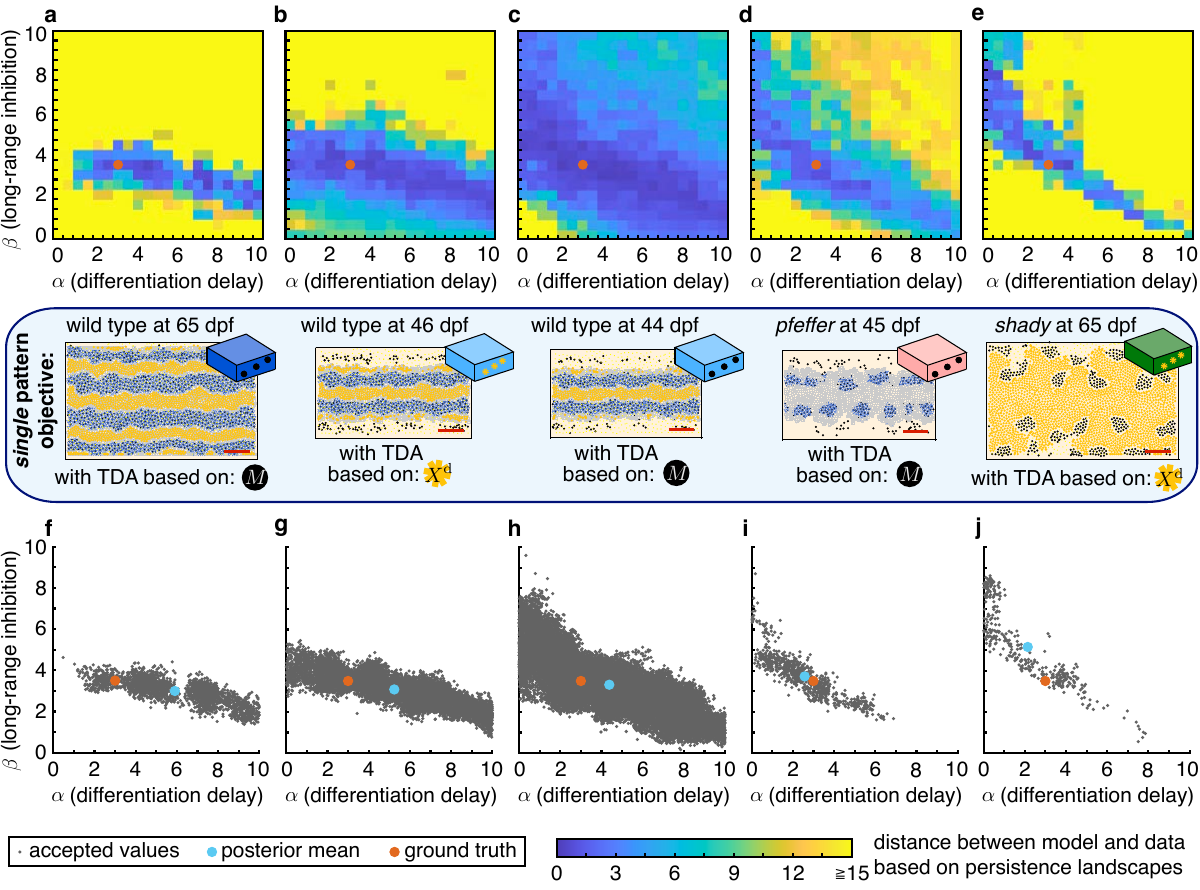} \vspace{-\baselineskip}
    \caption{Single objectives and non-identifiable parameters. Each column presents our inference results for $(\alpha, \beta)$ based on the single objective in the blue row. We show (a)--(e) heatmaps of the distance $d$ between model output and data (i.e., distance between mean persistence landscapes) and (f)--(j) the parameter values that we accept through AABC under each single objective. For example, (a, f) fitting to wild-type patterns at the final simulation time ($65$~dpf) is insufficient to achieve practical identifiability; while $\beta$ is identifiable, single-objective inference cannot distinguish between a broad range of $\alpha$ values. Inference based on (d, i) \textit{pfeffer} and (e, j) \textit{shady} under single objectives leads to different identifiability issues. (\textit{Nacre} is unsuitable for inference of $M$-differentiation parameters because it lacks $M$ cells.)  In (f)--(j), we apply the same threshold, accepting parameter proposals if $d < 1.2$; see Supporting Information~\ref{apx:fig_detail}.}  
    \label{fig:m-birth-params}
\end{figure*}

\subsection{Non-identifiability under single objectives}

As our first case study, we infer two parameters that govern $M$ proliferation. Based in the biological literature \cite{PatPLos,nakamasu2009InteractionsZebrafishPigment,Mahalwar,Budi,Dooley}, the ABM \cite{volkeningIridophoresSourceRobustness2018} describes $M$ differentiating from precursors due to long-range signals (Fig. \ref{fig:abm_schematic}(c)). Each day of simulated time, a number of randomly distributed precursors are evaluated for differentiation (if not overcrowded). The precursor at position $\textbf{x}$ becomes an $M$ cell if
\begin{align*}
\underbrace{\mathcal{N}( X^\text{d}, \Omega_\text{long}^\textbf{x})}_\text{$X^\text{d}$ at long range} + \underbrace{\mathcal{N}(I^\text{d}, \Omega_\text{long}^\textbf{x})}_\text{$I^\text{d}$ at long range} > \underbrace{\alpha}_\text{delay} + \underbrace{\beta \mathcal{N}(M, \Omega_\text{long}^\textbf{x})}_\text{$M$ at long range}, 
\end{align*}
where $\Omega_\text{long}^\textbf{x}$ is an annulus centered at $\textbf{x}$, and $\mathcal{N}(C, R)$ is the number of cells of type $C$ in the region $R$. This condition expresses that $X^\text{d}$ and $I^\text{d}$ promote $M$ differentiation at long range, and $M$ cells inhibit it \cite{nakamasu2009InteractionsZebrafishPigment,volkeningIridophoresSourceRobustness2018,volkening2015,PatPLos}. The parameter $\alpha \ge 0$ is a delay term, enforcing that there are more than $\alpha$ $X^\text{d}$ and $I^\text{d}$ cells present before $M$ appear, while $\beta \ge 0$ determines the strength of $M$ inhibition.

We first infer $\btheta = (\alpha, \beta)$ 
with various single objectives. Noting that $\alpha =3$~cells and $\beta = 3.5$ are the ground-truth values, we specify uniform priors with
\begin{align}
0 \leq \alpha \leq 10~\text{cells}, ~~~&~~~ 0 \leq \beta \leq 10 \text{ (unitless)}.
\label{eqn:m_prior}
\end{align}
As we show in Fig.~\ref{fig:m-birth-params} for a selection of single objectives, no one perspective on zebrafish results in a tight posterior distribution. For example, while our methods can determine $\beta$ fairly well in Fig.~\ref{fig:m-birth-params}(a), inference based only on wild type at the final simulation time is insufficient to identify $\alpha$. We conclude that the parameters $(\alpha, \beta)$ are practically non-identifiable when viewed from the perspective of a single experiment at one time, summarized through TDA. This matches our intuition, as the authors \cite{volkeningIridophoresSourceRobustness2018} required qualitative consistency across multiple experiments to specify parameters and model rules.

\subsection{Balancing model complexity with multiple objectives}

\begin{figure}
    \centering
    \includegraphics[width=0.5\textwidth]{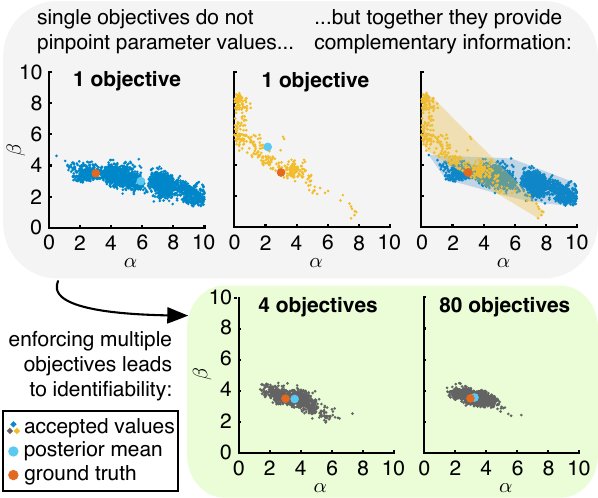}
    \caption{Combining multiple objectives to pinpoint parameters. Figure~\ref{fig:m-birth-params} shows that single objectives do not lead to practical identifiability, but the posterior distributions cover complementary regions of parameter space. In this example, the blue and yellow points are the posterior samples from Fig.~\ref{fig:m-birth-params}(f) and Fig.~\ref{fig:m-birth-params}(j), respectively. The intersection of their convex hulls (shaded regions) is much smaller than either convex hull individually, and it contains the ground-truth parameter values. Whether we combine just four objectives or eighty objectives together, we obtain practical identifiability of $(\alpha, \beta)$. The selection of these scores is based solely on rough knowledge of the pattern formation process; see Supporting Information~\ref{apx:fig_detail}.
    }
    \label{fig:posterior-intersection}
\end{figure}

When viewed individually, no posterior in Fig.~\ref{fig:m-birth-params} sufficiently constrains $(\alpha, \beta)$, yet the points in panels (f)--(g) and (i)--(j) cover different regions of parameter space. This suggests that using multiple objectives may better balance model complexity. We show this in Fig.~\ref{fig:posterior-intersection}: by applying our pipeline to $\btheta = (\alpha, \beta)$ with just four objectives, we achieve practical identifiability. Moreover, expanding to $80$ objectives leads to a similar posterior. 
Thus, a few carefully chosen objectives can provide enough information to infer parameters, but a broad swath of objectives can also work.

As a second test, we infer two cell-migration parameters. Each cell moves according to an ODE in the ABM \cite{volkeningIridophoresSourceRobustness2018}, so this example complements our first case study of stochastic rules. Specifically, the model \cite{volkeningIridophoresSourceRobustness2018} describes the movement of the $i$th $I^\text{d}$ cell, with position $\textbf{I}^\text{d}_i(t) \in \mathbb{R}^2$, as
\begin{align*}
    \frac{d\textbf{I}^\text{d}_i}{dt} &= \sum_{j=1, j \neq i}^{N^\text{d}} \textbf{F}^{\text{dd}}(\textbf{I}^\text{d}_i,\textbf{I}^\text{d}_j) 
    + \sum_{j=1}^{N^\text{d}} \textbf{F}^{\text{ld}}(\textbf{I}^\text{d}_i,\textbf{I}^\text{l}_j),
\end{align*}
where $\textbf{I}^\text{l}_j(t)$ is the position of the $j$th loose iridophore, $N^\text{d}$ and $N^\text{l}$ are the numbers of $I^\text{d}$ and $I^\text{l}$ cells, and
\begin{align*}
    \textbf{F}^{\mu \nu}(\textbf{x},\textbf{y}) &= R^{\mu \nu}\left(\frac{1}{2} + \frac{1}{2}\tanh{\frac{r_{\mu \nu} - \|\textbf{x} -\textbf{y}\|}{\delta}}\right) \frac{\textbf{x}-\textbf{y}}{\|\textbf{x}-\textbf{y}\|},
\end{align*}
with parameters $\delta$, $r_{\mu \nu}$, and $R^{\mu \nu}$; $I^\text{l}$ follow a similar ODE. We focus on the parameters $(R^\text{dd},R^\text{ll})$, which describe repulsion strength and have baseline values of $10$~micrometers/day ($\mu$m/day) and $32$~$\mu$m/day, respectively \cite{volkeningIridophoresSourceRobustness2018}.

\begin{figure*}[t!]
    \centering
    \includegraphics[width=\textwidth]{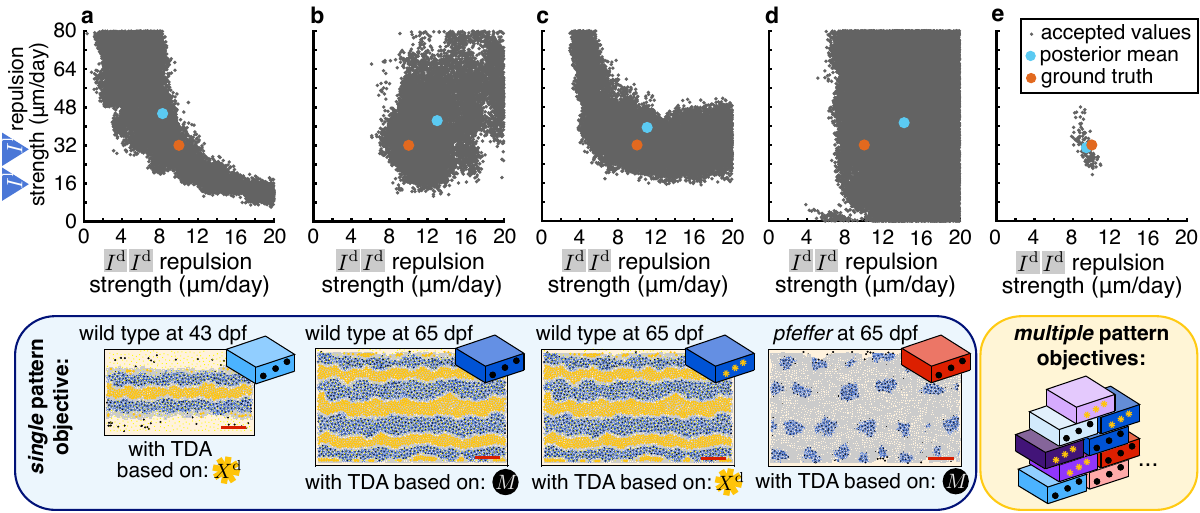} \vspace{-\baselineskip}
    \caption{Comparing single and multi-objective inference of migration parameters. To complement our study of stochastic rules for cell proliferation in Fig.~\ref{fig:m-birth-params}--Fig.~\ref{fig:posterior-intersection}, we illustrate the flexibility of our pipeline by applying it to two parameters in the differential equations governing iridophore movement. (a)--(d) A representative selection of single objectives shows that the repulsion strengths $(R^\text{dd},R^\text{ll})$ are not identifiable under a limited perspective on our data. (e) Inference based on a weighted sum of multiple objectives tightly and accurately constrains parameter values. We use $29$ objectives, consisting of wild type, \textit{pfeffer}, and \textit{nacre} at various timepoints; see Supporting Information~\ref{apx:fig_detail}.}
    \label{fig:movement_params}
\end{figure*}

Under a uniform prior distribution with \begin{align} 0 \le R^\text{dd} \le 20~\mu\text{m/day},~&~0 \le R^\text{ll} \le 80~\mu\text{m/day},
\label{eqn:mov_prior}
\end{align}
we infer $\btheta=(R^\text{dd},R^\text{ll})$ in Fig.~\ref{fig:movement_params}. Parallel to our study of $M$ differentiation, the results again show that---while any single objective may not constrain $\btheta$---requiring consistency across multiple objectives leads to a tight posterior with an accurate mean.

\subsection{Towards data-driven discovery of unknown cell-interaction rules}

Our first two case studies focus on parameter inference in ``known" ABM rules and equations. When faced with biological unknowns, however, a more fundamental question is whether alternative models are consistent with our data. (We distinguish between \textit{one model} with different parameter values, versus \textit{two models} with different rule structures.) For population-level models like $\textbf{u}_t = \mathcal{F}(x,t,\textbf{u},\textbf{u}_x,\dots)$, sparse regression approaches to equation learning \cite{brunton2016discovering,mangan2017model} provide a means of selecting from candidate terms in $\mathcal{F}(\cdot)$, such as a reaction that depends only on $u_1$ and a reaction that depends on two densities (e.g., $u_1u_2^2$). While powerful, this framework does not translate to rules in ABMs, which often require manual tuning and face long-standing questions about uniqueness. We expect that our methodology can help address this challenge, and, as a step in this direction, we show one way that our approach moves beyond parameter inference toward \textit{rule inference}, identifying a simpler model.

In our last case study, we investigate the rules that govern cell-type transitions between $I^\text{d}$ and $I^\text{l}$ in the ABM \cite{volkeningIridophoresSourceRobustness2018}. Unlike our earlier examples, at the time that the model \cite{volkeningIridophoresSourceRobustness2018} was published, nearly nothing was known about what cells iridophores communicate with during type transitions or where these cells are located. As one of their main contributions, the authors \cite{volkeningIridophoresSourceRobustness2018} proposed that an $I^\text{d}$ at position $\textbf{x}$ swaps to the $I^\text{l}$ form if
\begin{align}
&\underbrace{\mathcal{N}(M, \Omega_\text{loc}^\textbf{x}) > a}_{[\text{A}]\text{: enough $M$ locally}} ~~\textbf{or}~~\nonumber \\
&\left(\underbrace{\mathcal{N}(X^\text{d}, \Omega_\text{long}^\textbf{x})>b}_{[\text{B}]\text{: enough $X^\text{d}$ at long range}} \textbf{and}~ \underbrace{\mathcal{N}(X^\text{d}, \Omega_\text{loc}^\textbf{x})<c}_{[\text{C}]\text{: few $X^\text{d}$ locally}}\right), \label{eqn:irid_dense_loose}
\end{align}
and, inversely, an $I^\text{l}$ cell at $\textbf{x}$ changes to the $I^\text{d}$ form if
\begin{align}
&\underbrace{\mathcal{N}(M, \Omega_\text{loc}^\textbf{x}) < \tilde{a}}_{[\tilde{\text{A}}]\text{: few $M$ locally}} ~~\textbf{and}~~\nonumber \\ 
&\left( \underbrace{\mathcal{N}(X^\text{d}, \Omega_\text{long}^\textbf{x}) < \tilde{b}}_{[\tilde{\text{B}}]\text{: few $X^\text{d}$ at long range}} ~~\textbf{or}~~ \underbrace{\mathcal{N}(X^\text{d}, \Omega_\text{loc}^\textbf{x})>\tilde{c}}_{[\tilde{\text{C}}]\text{: enough $X^\text{d}$ locally}} \right), \label{eqn:irid_loose_dense}
\end{align}
where $\Omega_\text{long}^\textbf{x}$ is a long-range annulus neighborhood centered at $\textbf{x}$, $\Omega_\text{loc}^\textbf{x}$ is local ball neighborhood, and these rules are evaluated each simulated day; see ref.~\cite{volkeningIridophoresSourceRobustness2018} for details.

Equations \ref{eqn:irid_dense_loose} and \ref{eqn:irid_loose_dense} can be expressed as $[\text{A} \| (\text{B} \& \text{C})]$ and $[\tilde{\text{A}}\&(\tilde{\text{B}} \| \tilde{\text{C}})]$, respectively, where ``$\&$" enforces that two conditions must be met simultaneously and ``$\|$" indicates that at least one of two must be met. These nonlinear rules involve six parameters $\{a,b,c,\tilde{a},\tilde{b},\tilde{c}\}$, all with units of cell counts in specific regions. We make use of this feature to shift toward a rule-based inference perspective. Because $c$, $\tilde{a}$, and $\tilde{b}$ are upper bounds, setting one of them to zero means that its associated condition will never be satisfied. Alternatively, $a$, $b$, and $\tilde{c}$ are lower bounds, so we can turn their associated rules ``off" by setting these parameters very high. Together with the logical operators in \eqref{eqn:irid_dense_loose}--\eqref{eqn:irid_loose_dense} and the fact that all six parameters denote numbers of cells, this suggests that we can toggle the structure of our cell-interaction rules by changing these parameters to extreme values. For example, setting $a$ to be effectively infinite means that $[\text{A}]$ cannot be met, reducing Eq.~\ref{eqn:irid_dense_loose} from $[\text{A} \| (\text{B} \& \text{C})]$ to $[\text{B} \& \text{C}]$. If $b < 0$ as well, the number of $X^\text{d}$ cells in $\Omega_\text{long}$ is always greater than $b$, and Eq.~\ref{eqn:irid_dense_loose} reduces further to simply $[\text{C}]$.

\begin{figure*}[t!]
    \centering
    \includegraphics[width=\textwidth]{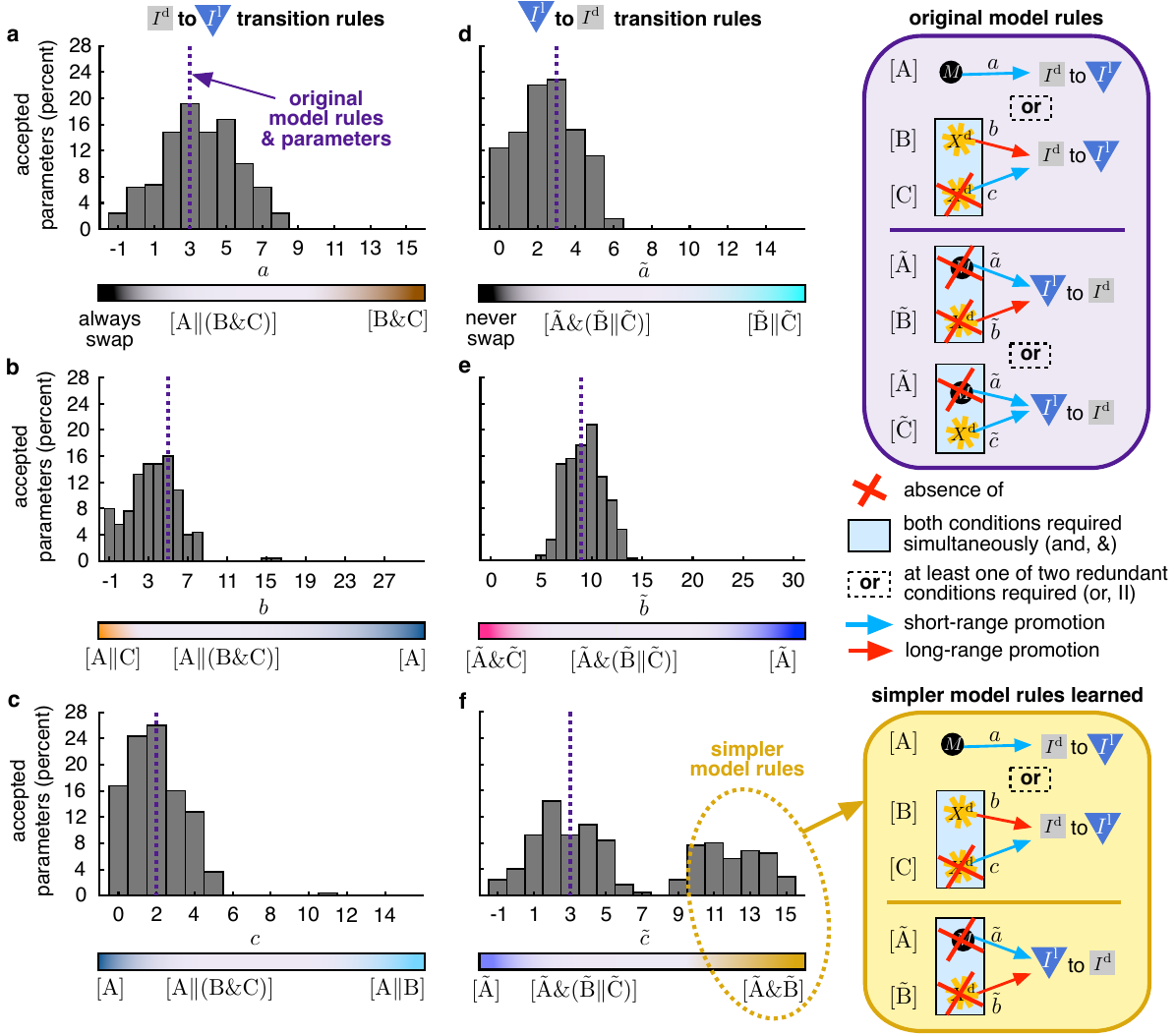} \vspace{-\baselineskip}
    \caption{Moving beyond parameter inference toward rule inference in ABMs. Here we infer the six parameters---$a$, $b$, $c$, $\tilde{a}$, $\tilde{b}$, and $\tilde{c}$, all with units of cells in local or long-range neighborhoods---involved in iridophore transitions in the model \cite{volkeningIridophoresSourceRobustness2018}. We present marginal posteriors for (a)--(c) the parameters associated with $I^\text{d}$-to-$I^\text{l}$ transitions and (d)--(f) the parameters governing $I^\text{l}$-to-$I^\text{d}$ transitions. Our prior distributions are so broad that values at their limits (e.g., $a= -1$ cells or $\tilde{c} =15$ cells) effectively turn off model rules and change the structure of cell interactions, allowing us to reframe parameter inference as rule learning across $81$ candidate models. The single-peak distributions in (a)--(e) indicate that these parameters are fairly tightly constrained, lending support to the original ABM \cite{volkeningIridophoresSourceRobustness2018}. However, the bimodal distribution in (f) highlights that an alternative, simpler model is consistent with our multi-experiment data. }
    \label{fig:irid-params-vary-acceptance}
\end{figure*}

With this in mind, we use our methodology to infer $\btheta =(a,b,c,\tilde{a},\tilde{b},\tilde{c})$ under an integer-valued prior given by
\begin{align}
-1 \leq a \leq 15, \quad -1 &\leq b \leq 30, \quad 0 \leq c \leq 15 \text{ cells} \nonumber \\
    0 \leq \tilde{a} \leq 15, \quad 0 &\leq \tilde{b} \leq 30, \quad -1 \leq \tilde{c} \leq 15 \text{ cells},\label{eqn:irid_params_prior}
\end{align}
where the upper bounds are high enough that they are effectively infinite for the cell densities in the ABM \cite{volkeningIridophoresSourceRobustness2018}. By setting $a$, $b$, and/or $c$ to be very high or low, our prior distribution contains nine candidate $I^\text{d}$-to-$I^\text{l}$ transition rules: $[\text{A} \| (\text{B} \& \text{C})]$, $[\text{B} \& \text{C}]$, $[\text{A} \| \text{B}]$, $[\text{A} \| \text{C}]$, $[\text{A}]$, $[\text{B}]$, $[\text{C}]$, $[\text{always swap to $I^\text{l}$}]$, and $[\text{never swap to $I^\text{l}$}]$. Similarly, nine $I^\text{l}$-to-$I^\text{d}$ rules are realizable in our prior. This means that applying our pipeline to infer $\btheta$ can be thought of as performing model selection across $81$ candidate models. Given the size of this task, we do so using $192$ objectives (i.e., all of our objectives, except early times and \textit{shady}, which lacks $I^\text{d}$ and $I^\text{l}$; see Supporting Information~\ref{apx:fig_detail}). The resulting marginal posteriors are in Fig.~\ref{fig:irid-params-vary-acceptance}. From a parameter-inference perspective,  the posteriors for $a$, $c$, $\tilde{a}$, and $\tilde{b}$ are unimodal and roughly centered at the ground-truth parameter values, suggesting practical identifiability. While the $b$ posterior includes a few moderate values, the marginal posterior for $\tilde{c}$ is clearly double-peaked.

More interestingly, we can also interpret Fig.~\ref{fig:irid-params-vary-acceptance} from a rule-inference perspective. Here the question becomes whether or not the posterior contains extreme $\btheta$ values that change the model structure. This corresponds to negative $a$, $b$, or $\tilde{c}$; zero values of $c$, $\tilde{a}$, or $\tilde{b}$; or values of any parameter at the upper limits of the prior in \eqref{eqn:irid_params_prior}. We see five such cases: first, the $c$ and $\tilde{a}$ marginals contain zero in their support, raising $[\text{A}]$ and $[\text{never swap to $I^\text{d}$}]$ as alternative rules. Moreover, the $\tilde{c}$ marginal distribution contains the lower and upper limits in its support, proposing $[\tilde{\text{A}}]$ and $[\tilde{\text{A}} \& \tilde{\text{B}}]$, respectively. (We note that individual marginals do not provide a full picture; 
for example, $b = -1$ in Fig.~\ref{fig:irid-params-vary-acceptance}(b) suggests $[\text{A} \| \text{C}]$, but closer inspection shows that the associated $\btheta$ values also have $c=0$, so the rule simplifies to $[\text{A}]$ again; see Supporting Information~\ref{apx:rule_comb}.) Lastly, the $a$ marginal contains $-1$, raising $[\text{always swap to $I^\text{l}$}]$. We conclude that our methodology suggests the original model in \eqref{eqn:irid_dense_loose}--\eqref{eqn:irid_loose_dense}, as well as alternative models with rule $[\text{A}]$ or $[\text{always swap to $I^\text{l}$}]$ for $I^\text{d}$ transitions, or rules $[\text{never swap to $I^\text{d}$}]$, $[\tilde{\text{A}}]$, or $[\tilde{\text{A}} \& \tilde{\text{B}}]$ for $I^\text{l}$ transitions\footnote{Because empirical understanding of iridophores has evolved \cite{gur} since ref.\ \cite{volkeningIridophoresSourceRobustness2018} was published, our results should be interpreted not as a biological prediction, but as a concrete example of rule inference in a detailed ABM built to address current unknowns at its time of publication.}. 

The rules $[\text{A}]$, $[\tilde{\text{A}}]$, $[\text{always swap to $I^\text{l}$}]$, and $[\text{never swap to $I^\text{d}$}]$ are inconsistent with the data that was available at the time of model publication \cite{volkeningIridophoresSourceRobustness2018}, and this points to a key final step. Because AABC involves simulating the ABM at a sparse set of parameter values, it relies on simulations from nearby parameter points as surrogates \cite{buzbasAABCApproximateApproximate2015}. This approximation smooths the estimated posterior distribution, consequently limiting its resolution and broadening its support. As a result, we suggest that parameter values at the boundaries of a peak may be unreliable. To address this, after identifying model candidates with our approach, the last step is a posterior predictive check: we evaluate $500$ parameters sampled from the distribution in Fig.~\ref{fig:irid-params-vary-acceptance} for qualitative agreement with wild type, \textit{pfeffer}, and \textit{nacre} at the final simulation time, classifying them as ``good" if they perform well on all three patterns. We find that low $a$, $c$, $\tilde{a}$, and $\tilde{c}$ do not perform well, indicating that these edge cases represent blurring through AABC rather than true alternative models. Moreover, only about $13\%$ of our $500$ parameter samples perform well based on rough manual evaluation. This observation further illustrates that the cost of performing a computationally feasible number of ABM simulations is sharpness in the posteriors. Critically, however, we find that some values in the second peak of the $\tilde{c}$ distribution lead to good patterns across our experiments (Fig.\ S2). We conclude that $[\tilde{\text{A}} \& \tilde{\text{B}}]$ is a true alternative to model rule $[\tilde{\text{A}} \& (\tilde{\text{B}} \| \tilde{\text{C}}]$. In the future, applying our methodology to identify candidate rules and then inferring their parameters under a much reduced prior may be a practical way to refine posteriors.

As a whole, these results confirm that our multi-objective pipeline is able to identify redundancy in cell interactions and suggest alternative, simpler ABM rules. We see our approach as involving broad conceptual parallels to regression-based equation learning \cite{brunton2016discovering,mangan2017model}, in the sense that we establish candidate cell-interaction rules and then infer both model structure and parameter values at the same time. While some challenges remain, our pipeline makes it possible to consider $81$ candidate rules in our last case study---an infeasible task for a modeler to complete by hand, especially when it is necessary to view multiple experiments, times, and stochastic replicates. Additionally, by searching across those $81$ candidates, we find just two models clearly supported by our data.

\section{Discussion}\label{sec:discussion}

Stochastic ABMs can make powerful predictions about spatial pattern formation, but they also face long-standing questions about parameter identifiability and model uniqueness. To help address this challenge, we developed a pipeline for inferring parameters and conducting practical identifiability analysis in detailed ABMs. Our approach combines TDA and AABC, balancing model complexity with consistency across multiple experiments. Persistent homology provides a means of quantifying spatial data, and AABC \cite{buzbasAABCApproximateApproximate2015,Lambert2018} allows us to perform Bayesian inference while keeping computational cost affordable.

We demonstrated our methodology by applying it to a biologically detailed ABM \cite{volkeningIridophoresSourceRobustness2018} for cell behavior during zebrafish pattern formation. Our first two examples, involving a stochastic rule for cell proliferation and ODEs for migration, illustrated the value of a multi-objective perspective. We found that multiple experiments were necessary to obtain accurate estimates of two parameters and achieve identifiability. We then turned to the more difficult task of inferring six parameters governing poorly understood cell behaviors \cite{volkeningIridophoresSourceRobustness2018}. By introducing prior distributions with extreme values, we reframed parameter inference as a search across candidate rules and were able to identify a simpler model consistent with our data. This example illustrated the promise of our approach as a means of moving beyond parameter inference in known model rules, toward inference of the rules themselves.

In terms of building on our approach, developing methods for selecting AABC hyperparameters is a valuable future direction. The acceptance threshold controls how many parameter values are accepted into the posterior distribution, and it is common for studies \cite{Lambert2018,wenzel2025TopologicallybasedParameterInference,thorneTopologicalApproximateBayesian2022} that use ABC or AABC to set it in a largely \textit{ad hoc} way, as we did here. However, this threshold can affect qualitative features of the posterior. For example, if we reduce the acceptance threshold, all of the marginals in Fig.~\ref{fig:m-birth-params} become unimodal (Fig.\ S3) and we lose the second peak that signaled us to the presence of an alternative model. Moreover, sparse sampling may fail to capture finer features of the likelihood landscape, which are smoothed out by the interpolating nature of AABC. We suggest that one feasible way to mitigate this limitation in future work is to perform multiple iterations of AABC, using the posterior from one round as the prior for the next round.

The presence of multiple objectives was key to our work, and, in the case of the ABM \cite{volkeningIridophoresSourceRobustness2018}, these objectives took the form of wild-type and mutant patterns. In the future, it will be interesting to apply our approach to other ABMs that involve different experiments, such as altered initial conditions in scratch assays \cite{Perez2022}. It will also be useful to consider alternative methods for combining and weighting objectives. Through these and other future directions, we expect that the combination of rich spatial data, detailed methods for quantifying patterns, mechanistic modeling, and data-driven approaches will continue to increase the tractability and predictive power of agent-based models.

\section{Methods}

\subsection{Generating ground-truth data}

We estimate parameters based on \textit{in silico} data so that our parameter values are known, providing a benchmark for our methodology. We simulate the ABM \cite{volkeningIridophoresSourceRobustness2018} five times per experiment under its baseline parameters, leading to five daily timelines from $21$ to $65$ dpf for each condition (wild type, \textit{pfeffer}, \textit{nacre}, and \textit{shady}). As in ref. \cite{volkeningIridophoresSourceRobustness2018}, our ground-truth simulations have $\alpha = 3$~cells, $\beta = 3.5$ (unitless), $R^\text{dd} = 10$~$\mu$m/day, $R^\text{ll} = 32$~$\mu$m/day, $a= 3$~cells, $b=5$~cells, $c=2$~cells, $\tilde{a} = 3$~cells, $\tilde{b} = 9$~cells, and $\tilde{c} = 3$~cells. We use $10001, \dots, 10005$ as seeds (chosen arbitrarily) with the Mersenne Twister algorithm.

\subsection{Computing persistent homology}

For each pattern, we compute persistent homology based on the $(x,y)$-coordinates of the selected cell type at the given time point. Following prior TDA studies \cite{mcguirlTopologicalDataAnalysis2020,clevelandQuantifyingDifferentModeling2023} of zebrafish, we first remove any cells in the top or bottom $10$\% of the domain. (Because patterning progresses from the center of the domain outward, cropping limits noise due to random differentiation and focuses attention on the most complete part of the pattern \cite{volkeningIridophoresSourceRobustness2018}.) Using the remaining cells as vertices, we implement the Vietoris--Rips filtration for dimensions $0$ and $1$ in \textit{Ripser} \cite{bauer2021RipserEfficientComputation} to construct a family of simplicial complexes indexed by the filtration scale $\varepsilon \le \varepsilon_\text{max} = 650~\mu\text{m}$ (Fig.~\ref{fig:tda_schematic}(b)).
Because the ABM \cite{volkeningIridophoresSourceRobustness2018} enforces periodic boundaries horizontally, we do so when computing persistent homology.

\subsection{Constructing persistence landscapes}

To compute a persistence landscape, one first rotates its persistence diagram so that the $\varepsilon_\text{birth} = \varepsilon_\text{death}$ line is the horizontal axis (Fig.~\ref{fig:tda_schematic}(d)). This means setting $r = (\varepsilon_\text{birth}+\varepsilon_\text{death})/2$ and $p = (\varepsilon_\text{death}-\varepsilon_\text{birth})/2$ and plotting the point $(r,p)$ for each feature in the chosen dimension. Next imagine that each point $(r,p)$ marks the tip of a right triangle with two of its sides extending from that point down toward the $r$-axis, and its third side along the $r$-axis. Then $\lambda_\kappa$ is the envelope of the region in the $(r,p)$-plane covered by at least $\kappa$ triangles (Fig.~\ref{fig:tda_schematic}(e)). The intuition behind persistence landscapes is that features with short persistence---often considered to be less significant---appear near the horizontal $r$-axis, contributing little to the landscape curves. In this way, each feature's contribution depends on its persistence. More specifically, a persistence landscape \cite{bubenik2015StatisticalTopologicalData} is a family of functions $\lambda_\kappa(r)$, where $\lambda_\kappa(r)$ is the $k$th largest value of $f(r,\varepsilon_\text{birth}^i,\varepsilon_\text{death}^i)$. Here $f(r,x,y) = \min(r-x,y-r)_+$,
$\min(a,b)_+ = 0$ if $\min(a,b) < 0$ and $\min(a,b)_+ = \min(a,b)$ otherwise, and $\{(\varepsilon_\text{birth}^i,\varepsilon_\text{death}^i)\}$ is the set of points in the persistence diagram for the chosen dimension. To construct persistence landscapes, we use the \textit{PersLandApprox} routine in the \textit{Persim} package \cite{scikittda2019}. This routine approximates the landscape by sampling the curves at $r = r_1, r_2, \dots, r_{N_p}$ \cite{bubenik2017PersistenceLandscapesToolbox}. We use $N_p = 600$ with $r_1=0$ and a grid step of $600/599$~$\mu$m, and compute the first $N_\kappa = 300$ landscape curves. As $\kappa$ grows, these curves are more likely to be zero, and the information that they contain becomes less meaningful for our application.

\subsection{Quantifying pattern distance under multiple objectives}

Persistence landscapes are amenable to averaging \cite{Hartsock2025,bubenik2015StatisticalTopologicalData}, and we make use of this in our inference pipeline. To account for biological variability and model stochasticity, each parameter proposal $\btheta$ is associated with $N_\text{rep} =5$ replicates for each pattern of interest. ``Pattern of interest" refers to a given experiment at a given time. We also select the cell type whose positions will serve as vertices when computing persistent homology and the dimension that we will focus on when building persistence landscapes. Thus, for a given experiment $E \in \{\text{wild type}, \textit{pfeffer}, \textit{nacre}, \textit{shady}\}$ and time $T \in \{21,22,\dots,65\}$ dpf, quantitatively summarized from the perspective of cell type $C \in \{M,X^\text{d},X^\text{l}\}$ and topological dimension $D \in \{0,1\}$, we define the average persistence landscape curves associated with $\btheta$ as
\begin{align}\Lambda_\kappa(r_j; E, T, C, D) &= \frac{1}{N_\text{rep}}\sum_{\ell=1}^{N_\text{rep}} \lambda_\kappa^\ell(r_j; E, T, C, D), \label{eq:lambda}
\end{align}
where $\kappa = 1, 2,\dots, N_\kappa$ denotes the depth of the landscape, and $\lambda_\kappa^\ell$ is associated with the $\ell^\textrm{th}$ replicate. These mean landscape curves (across $5$ replicates) provide a quantitative summary of the patterns generated under parameter $\btheta$, viewed from the perspective of our chosen experiment, time, cell type, and topological dimension.

Given the mean persistence landscape $\{\Lambda_\kappa^\text{data}(r_j; E, T, C, D)\}_{\kappa=1}^{N_\kappa}$ for our ground-truth data, we measure the distance between the patterns associated with parameter proposal $\btheta$ and ground truth according to
\begin{align}\tilde{d}(\btheta; E, T, C, D) = \frac{1}{N_\kappa N_p} \sum_{\kappa=1}^{N_\kappa} \sum_{j=1}^{N_p} ( \Lambda_\kappa&(r_j; E, T, C, D) -  \nonumber \\
&\Lambda_\kappa^\text{data}(r_j; E, T, C, D))^2.\label{eq:singleD}
\end{align}
Here \eqref{eq:singleD} serves as a single objective, scoring $\btheta$ according to one set of pattern replicates for a single experiment at a single time point, quantified by applying persistent homology to a single cell type and constructing persistence landscapes in a single dimension. To account for multiple objectives, we instead consider the more flexible distance function
\begin{align}d(\btheta) &= \frac{1}{N_\text{obj}}\sum_{i=1}^{N_\text{obj}} w_i ~\tilde{d}(\btheta;~\text{perspective $i$}), \label{eq:multiD}
\end{align}
where perspective $i = (\text{exp.}~E_i, \text{time}~T_i, \text{cell type}~C_i, \text{dim.}~D_i)$ and $N_\text{obj}$ is the number of objectives. The weights $\{w_i\}$ control how strong a role each perspective plays. We set $w_i$ to be the inverse of the minimum score $\tilde{d}(\btheta^*_j; E_i, T_i, C_i, D_i)$ across the pre-proposals $\{\btheta^*_j\}$ in the first AABC step:
\begin{align}
    w_i &= \left(\min_\text{pre-proposals $j$} ~\tilde{d}(\btheta^*_j, \text{exp.}~E_i, \text{time}~T_i, \text{cell type}~C_i, \text{dim.}~D_i)\right)^{-1},
\end{align}
where $w_i$ is the weight that we give the $i$th quantitative perspective. (For example, perspective $i$ could correspond to looking at 
\textit{pfeffer} at $45$~dpf by constructing persistence landscapes in dimension~$0$ based on $M$ cells.)

\subsection{Inferring parameters through multi-objective AABC}

We estimate the posterior distribution of our parameters $\btheta$ through AABC \cite{buzbasAABCApproximateApproximate2015,Lambert2018}. Our implementation follows the framework of Buzbas and Rosenberg \cite{buzbasAABCApproximateApproximate2015}, 
with adaptions to account for multiple objectives. We think of AABC as a three-stage process: pre-proposal, pseudo-simulation, and acceptance.

We implement the pre-proposal stage of AABC as \textbf{Steps~(I)--(II)} in Fig.~\ref{fig:aabc_schematic} of the Supporting Information. First, we sample $N_\text{pre}$ parameter values $\{\btheta^*_j\}_{j=1,\dots,N_\text{pre}}$ from our prior. For each ``pre-proposal" $\btheta^*$, we simulate the ABM \cite{volkeningIridophoresSourceRobustness2018} $N_\text{rep}$ times under each experiment of interest and apply TDA to quantify the results. This means that, for each replicate of experiment $E$, we generate daily snapshots of \textit{in silico} patterning from $21$ to $65$ dpf. For each time point, depending on our choice of perspective, we build as many as six persistence landscapes per replicate, describing the shape of our pattern in dimension $0$ or $1$ based on $M$, $X^\text{d}$, or $X^\text{l}$. The result of the pre-proposal stage is $N_\text{pre} \times N_\text{rep} \times N_\text{obj}$ persistence landscapes, where one parameter sample $\btheta^*$ is responsible for generating each set of $N_\text{rep} \times N_\text{obj}$ landscapes. Here $N_\text{obj}$ is the number of objectives (determined based on our chosen perspectives $\{(E_i, T_i,C_i,D_i)\}_{i=1,\dots,N_\text{obj}}$, where experiment $E_i \in \{\text{wild type}, \textit{pfeffer},\textit{nacre},\textit{shady}\}$, time $T_i \in \{21, 22, \dots, 65\}$, cell type $C_i \in \{M, X^\text{d},X^\text{l}\}$ for persistent homology, and topological dimension $D_i \in \{0,1\}$ for constructing persistence landscapes). Based on the broad uniform priors described by \eqref{eqn:m_prior}--\eqref{eqn:mov_prior}, we specify pre-proposals as
$\{(\alpha^*,\beta^*)\} = \{(0.5i, 0.5j)\}_{i,j=0, \dots, 20}$ and $\{(R^{\text{dd}*},R^{\text{ll}*})\} = \{(i, 4j)\}_{i,j=0, \dots, 20}$. For iridophore transitions, we sample $4750$ pre-proposals from the prior defined in \eqref{eqn:irid_params_prior}. Because the true parameter values are included as grid sites for $\alpha$, $\beta$, $R^\text{dd}$, and $R^\text{ll}$, we add the true values, so we have $4751$ pre-proposals for $\btheta = (a,b,c,\tilde{a},\tilde{b},\tilde{c})$. In the pre-proposal stage of AABC, we simulate the ABM \cite{volkeningIridophoresSourceRobustness2018} five times per experiment under each $\btheta^*$, using random seeds $2024001, \dots, 2024005$ (chosen arbitrarily to be different than the seeds for ground-truth data).

The pseudo-simulation stage of AABC replaces the 
original model with a computationally efficient statistical model (\textbf{Step~(III)} in Fig.~\ref{fig:aabc_schematic}). First, we sample $N_\text{pseu} \gg N_\text{pre}$ parameter values $\{\btheta'_j\}_{j=1,\dots,N_\text{pseu}}$ from the prior. Second, instead of simulating the ABM \cite{volkeningIridophoresSourceRobustness2018} under these ``pseudo-proposals", we reuse the quantitative summaries that we generated from our pre-proposals in Step~(II). Broadly, we assign each pseudo-proposal $N_\text{rep} \times N_\text{obj}$ persistence landscapes by randomly selecting from those associated with nearby pre-proposals. For each pseudo-proposal, this process \cite{buzbasAABCApproximateApproximate2015} has five sub-steps. In \textbf{Step~III.1} we find the closest $N_\text{nei}=5$ pre-proposals to $\btheta'$, denoting $\btheta^{*}_{k\text{th close}}$ as the $k$th nearest to $\btheta'$. In \textbf{Step~III.2}, we compute weights $\alpha_j$ based on the Epanechnikov kernel \cite{buzbasAABCApproximateApproximate2015}:
        \[ \alpha_k = \frac{3}{4} \frac{1}{\|\btheta' -\btheta^*_{(N_\text{nei}+1)\text{th close}}\|}\left[1-\left(\frac{\|\btheta' -\btheta^*_{k\text{th close}}\|}{\|\btheta' -\btheta^*_{(N_\text{nei}+1)\text{th close}}\|} \right)^2 \right],\]
       where $k= 1, \dots, N_\text{nei}$. This nearest pre-proposal to $\btheta'$ receives the highest weight $\alpha_1$. In \textbf{Step III.3}, we determine the pre-proposals from whom $\btheta'$ will inherit data. Choosing from the $N_\text{nei}$ nearest pre-proposals to $\btheta'$, we sample (with replacement) from $\{\btheta^*_{k\text{th close}}\}_{k=1,\cdots,N_\text{nei}}$ to select a set $\{\btheta^*_{k_s}\}_{s=1,\dots,N_\text{nei}}$. As in ref.\ \cite{buzbasAABCApproximateApproximate2015}, this sampling depends on distance, with $p_k$, the probability of selecting the $k$th nearest pre-proposal to $\btheta'$ depending on the Dirichlet distribution with parameters $[\alpha_k]$. The result is that a set of pre-proposals $\{\btheta^*_{k_s}\}_{s=1,\dots,N_\text{nei}}$ is now associated with $\btheta'$. 
        
        In \textbf{Step III.4}, for each $\btheta^*_{k_s}$, we sample uniformly at random from its persistence landscapes to assign $\btheta'$ a set of $N_\text{obj}$ landscapes per objective. We enforce that the same replicates (i.e., simulations under specific random seeds) serve as the basis of all objectives. This ensures that we assign $\btheta'$ consistent trajectories (rather than a few time points from a simulation under $\btheta^*_{k_s}$, and a few from one under $\btheta^*_{k_s}$). In \textbf{Step III.5}, for each objective, we aggregate its $N_\text{obj}$ landscapes through \eqref{eq:lambda}. The pseudo-proposal stage leads to $N_\text{pseu} \times N_\text{obj}$ mean persistence landscapes. The set of mean landscape curves $\{\Lambda_\kappa(r_j; E_i, T_i, C_i, D_i)\}_{\kappa=1}^{N_\kappa}$ describes the average pattern dynamics under $\btheta'$ for perspective $(E_i, T_i, C_i, D_i)$. Lastly, in the acceptance stage of AABC, we construct our posterior (\textbf{Step~(IV)} in Fig.~\ref{fig:aabc_schematic}). Considering the mean landscapes for our ground-truth data and those associated with $\btheta'$, we accept $\btheta'$ into the posterior if $d(\btheta') < \delta$, and otherwise reject $\btheta'$; see \eqref{eq:singleD}--\eqref{eq:multiD}. Repeating this process leads to an estimate of the posterior for $\btheta$. We use $N_\text{pseu} = 10^5$ when we infer two parameters a time, and $N_\text{pseu} = 10^6$ when we infer six parameters.\\

\noindent \textbf{Code availability:} The code that we developed to implement our inference pipeline will be made available upon publication.\\

\noindent \textbf{Acknowledgments:} AV gratefully acknowledges Adam L.\ MacLean for suggesting AABC to her.

\printbibliography

\renewcommand{\thefigure}{S\arabic{figure}}
\setcounter{figure}{0}
\section{Supporting Information}

\subsection{Summary of Models Considered in our Rule Inference Case Study}\label{apx:rule_comb}

In our last case study in the main manuscript, we introduce extended prior distributions that contain extreme low and high values for the number of cells in different regions. Here we illustrate how changing the parameters $a,b,c,\tilde{a},\tilde{b}$, and $\tilde{c}$ in Eq.~(3)--Eq.~(4) in the main manuscript leads to $81$ candidate models. As a guide, we reproduce the relevant ABM \cite{volkeningIridophoresSourceRobustness2018} rules, which can be represented as $[\text{A} \| (\text{B} \& \text{C})]$ and $[\tilde{\text{A}} \& (\tilde{\text{B}} \| \tilde{\text{C}})]$, below:
\begin{align*}
\underbrace{\mathcal{N}(M, \Omega_\text{loc}^\textbf{x}) > a}_{[\text{A}]\text{: enough $M$ locally}} ~~\textbf{or}~~
\left(\underbrace{\mathcal{N}(X^\text{d}, \Omega_\text{long}^\textbf{x})>b}_{[\text{B}]\text{: enough $X^\text{d}$ at long range}} \textbf{and}~ \underbrace{\mathcal{N}(X^\text{d}, \Omega_\text{loc}^\textbf{x})<c}_{[\text{C}]\text{: few $X^\text{d}$ locally}}\right) ~ &\Longrightarrow ~ I^\text{d} \text{ at position } \textbf{x} \text{ swaps to } I^\text{l},\\
\underbrace{\mathcal{N}(M, \Omega_\text{loc}^\textbf{x}) < \tilde{a}}_{[\tilde{\text{A}}]\text{: few $M$ locally}} ~~\textbf{and}~~
\left( \underbrace{\mathcal{N}(X^\text{d}, \Omega_\text{long}^\textbf{x}) < \tilde{b}}_{[\tilde{\text{B}}]\text{: few $X^\text{d}$ at long range}} ~~\textbf{or}~~ \underbrace{\mathcal{N}(X^\text{d}, \Omega_\text{loc}^\textbf{x})>\tilde{c}}_{[\tilde{\text{C}}]\text{: enough $X^\text{d}$ locally}} \right) ~&\Longrightarrow ~ I^\text{l} \text{ at position } \textbf{x} \text{ swaps to } I^\text{d},
\end{align*}
where $\Omega^\textbf{x}_\text{long}$ is a long-range annulus centered at position $\textbf{x}$ and can be thought of as capturing interactions through long cell extensions (see Fig.~2(c) in the main manuscript), and $\Omega^\textbf{x}_\text{loc}$ accounts for local and short-range interactions in a ball neighborhood centered at $\textbf{x}$; see ref. \cite{volkeningIridophoresSourceRobustness2018} for full details and length scales. Importantly, the upper bounds of our prior in Eq.~(5) in the main manuscript can be considered to be effectively infinite from the perspective of the cell densities in the ABM \cite{volkeningIridophoresSourceRobustness2018}.

Considering $I^\text{d}$-to-$I^\text{l}$ transitions first, we notice that $a = -1$ means that $[\text{A}]$ is always satisfied, since the number of cells in any neighborhood is always at least zero. For the same reason, $b = -1$ means that $[\text{B}]$ is always satisfied. On the other hand, if $c$ is large (effectively infinite), condition $[\text{C}]$ is always satisfied. Moreover, if $a$ or $b$ are large (effectively infinite) it becomes practically infeasible to satisfy $[\text{A}]$ or $[\text{B}]$, respectively. Putting this together, we have:
\begin{itemize}
\item $a$ moderate, $b$ moderate, $c$ moderate: \textbf{$[\text{A} \| (\text{B} \& \text{C})]$ (original model rules)},
\item $a$ moderate, $b$ moderate, $c$ large:\textbf{ $[\text{A} \| \text{B}]$},
\item $a$ moderate, $b$ moderate, $c$ zero: \textbf{$[\text{A}]$},
\item $a$ moderate, $b$ large, $c$ moderate: $[\text{A}]$,
\item $a$ moderate, $b$ large, $c$ large: $[\text{A} \| \text{B}]$,
\item $a$ moderate, $b$ large, $c$ zero: $[\text{A}]$,
\item $a$ moderate, $b$ negative, $c$ moderate: \textbf{$[\text{A} \| \text{C}]$},
\item $a$ moderate, $b$ negative, $c$ large: \textbf{$[\text{always swap to $I^\text{l}$}]$},
\item $a$ moderate, $b$ negative, $c$ zero: $[\text{A}]$,
\item $a$ large, $b$ moderate, $c$ moderate: \textbf{$[\text{B} \& \text{C}]$}, 
\item $a$ large, $b$ moderate, $c$ large: \textbf{$[\text{B}]$},
\item $a$ large, $b$ moderate, $c$ zero: \textbf{$[\text{never swap to $I^\text{l}$}]$},
\item $a$ large, $b$ large, $c$ moderate: $[\text{never swap to $I^\text{l}$}]$,
\item $a$ large, $b$ large, $c$ large: $[\text{never swap to $I^\text{l}$}]$,
\item $a$ large, $b$ large, $c$ zero: $[\text{never swap to $I^\text{l}$}]$,
\item $a$ large, $b$ negative, $c$ moderate: \textbf{$[\text{C}]$},
\item $a$ large, $b$ negative, $c$ large: $[\text{always swap to $I^\text{l}$}]$,
\item $a$ large, $b$ negative, $c$ zero: $[\text{never swap to $I^\text{l}$}]$,
\item $a$ negative, $b$ moderate, $c$ moderate: $[\text{always swap to $I^\text{l}$}]$,
\item $a$ negative, $b$ moderate, $c$ large: $[\text{always swap to $I^\text{l}$}]$,
\item $a$ negative, $b$ moderate, $c$ zero: $[\text{always swap to $I^\text{l}$}]$,
\item $a$ negative, $b$ large, $c$ moderate: $[\text{always swap to $I^\text{l}$}]$,
\item $a$ negative, $b$ large, $c$ large: $[\text{always swap to $I^\text{l}$}]$,
\item $a$ negative, $b$ large, $c$ zero: $[\text{always swap to $I^\text{l}$}]$,
\item $a$ negative, $b$ negative, $c$ moderate: $[\text{always swap to $I^\text{l}$}]$,
\item $a$ negative, $b$ negative, $c$ large: $[\text{always swap to $I^\text{l}$}]$, and
\item $a$ negative, $b$ negative, $c$ zero: $[\text{always swap to $I^\text{l}$}]$,
\end{itemize}
where we \textbf{bold} the nine unique $I^\text{d}$-to-$I^\text{l}$ rules that we obtain by changing the parameters $a,b$, and $c$.

Similarly, we find the $I^\text{l}$-to-$I^\text{d}$ rules contained in our prior distribution by changing the parameters $\tilde{a}$, $\tilde{b}$, and $\tilde{c}$, as below:
\begin{itemize}
\item $\tilde{a}$ moderate, $\tilde{b}$ moderate, $\tilde{c}$ moderate: \textbf{$[\tilde{\text{A}} \& (\tilde{\text{B}} \| \tilde{\text{C}})]$ (original model rules)},
\item $\tilde{a}$ moderate, $\tilde{b}$ moderate, $\tilde{c}$ large:\textbf{$[\tilde{\text{A}} \& \tilde{\text{B}}]$},
\item $\tilde{a}$ moderate, $\tilde{b}$ moderate, $\tilde{c}$ negative: \textbf{$[\tilde{\text{A}}]$},
\item $\tilde{a}$ moderate, $\tilde{b}$ large, $\tilde{c}$ moderate: $[\tilde{\text{A}}]$,
\item $\tilde{a}$ moderate, $\tilde{b}$ large, $\tilde{c}$ large: $[\tilde{\text{A}}]$,
\item $\tilde{a}$ moderate, $\tilde{b}$ large, $\tilde{c}$ negative: $[\tilde{\text{A}}]$,
\item $\tilde{a}$ moderate, $\tilde{b}$ zero, $\tilde{c}$ moderate: \textbf{$[\tilde{\text{A}} \& \tilde{\text{C}}]$},
\item $\tilde{a}$ moderate, $\tilde{b}$ zero, $\tilde{c}$ large: \textbf{$[\text{never swap to $I^\text{d}$}]$},
\item $\tilde{a}$ moderate, $\tilde{b}$ zero, $\tilde{c}$ negative: $[\tilde{\text{A}}]$,
\item $\tilde{a}$ large, $\tilde{b}$ moderate, $\tilde{c}$ moderate: \textbf{$[\tilde{\text{B}} \| \tilde{\text{C}}]$}, 
\item $\tilde{a}$ large, $\tilde{b}$ moderate, $\tilde{c}$ large: \textbf{$[\tilde{\text{B}}]$},
\item $\tilde{a}$ large, $\tilde{b}$ moderate, $\tilde{c}$ negative: \textbf{$[\text{always swap to $I^\text{d}$}]$},
\item $\tilde{a}$ large, $\tilde{b}$ large, $\tilde{c}$ moderate: $[\text{always swap to $I^\text{d}$}]$,
\item $\tilde{a}$ large, $\tilde{b}$ large, $\tilde{c}$ large: $[\text{always swap to $I^\text{d}$}]$,
\item $\tilde{a}$ large, $\tilde{b}$ large, $\tilde{c}$ negative: $[\text{always swap to $I^\text{d}$}]$,
\item $\tilde{a}$ large, $\tilde{b}$ zero, $\tilde{c}$ moderate: \textbf{$[\tilde{\text{C}}]$},
\item $\tilde{a}$ large, $\tilde{b}$ zero, $\tilde{c}$ large: $[\text{never swap to $I^\text{d}$}]$,
\item $\tilde{a}$ large, $\tilde{b}$ zero, $\tilde{c}$ negative: $[\text{always swap to $I^\text{d}$}]$,
\item $\tilde{a}$ zero, $\tilde{b}$ moderate, $\tilde{c}$ moderate: $[\text{never swap to $I^\text{d}$}]$,
\item $\tilde{a}$ zero, $\tilde{b}$ moderate, $\tilde{c}$ large: $[\text{never swap to $I^\text{d}$}]$,
\item $\tilde{a}$ zero, $\tilde{b}$ moderate, $\tilde{c}$ negative: $[\text{never swap to $I^\text{d}$}]$,
\item $\tilde{a}$ zero, $\tilde{b}$ large, $\tilde{c}$ moderate: $[\text{never swap to $I^\text{d}$}]$,
\item $\tilde{a}$ zero, $\tilde{b}$ large, $\tilde{c}$ large: $[\text{never swap to $I^\text{d}$}]$,
\item $\tilde{a}$ zero, $\tilde{b}$ large, $\tilde{c}$ negative: $[\text{never swap to $I^\text{d}$}]$,
\item $\tilde{a}$ zero, $\tilde{b}$ zero, $\tilde{c}$ moderate: $[\text{never swap to $I^\text{d}$}]$,
\item $\tilde{a}$ zero, $\tilde{b}$ zero, $\tilde{c}$ large: $[\text{never swap to $I^\text{d}$}]$, and
\item $\tilde{a}$ zero, $\tilde{b}$ zero, $\tilde{c}$ negative: $[\text{never swap to $I^\text{d}$}]$,
\end{itemize}
where we \textbf{bold} the nine unique $I^\text{l}$-to-$I^\text{d}$ rules present. Taken together with our nine rules for $I^\text{d}$-to-$I^\text{l}$ transitions, we conclude that toggling the parameters $a,b,c,\tilde{a},\tilde{b}$, and $\tilde{c}$ leads to $81$ models.

\subsection{Reproducibility Information for Figures}\label{apx:fig_detail}

Here we provide further details on how to reproduce the results in all of our case studies presented as figures; also see Methods in the main manuscript. For each figure, we consider a specific set of objectives $\{(E_i,T_i,C_i,D_i)\}_{i=1,\dots,N_\text{obj}}$, where experiment $E_i \in \{\text{wild type}, \textit{pfeffer}, \textit{nacre}, \textit{shady}\}$, developmental time $T_i \in \{21,22,\dots,65\}$ days post fertilization (dpf), cell type $C_i \in \{M,X^\text{d},X^\text{l}\}$, and topological dimension $D_i \in \{0,1\}$. In addition to selecting objectives, our pipeline includes two key hyperparameters: the number $N_\text{pseu}$ of pseudo-simulations (i.e., parameter samples from the prior) that we generate through our statistical surrogate model as part of AABC \cite{buzbasAABCApproximateApproximate2015}, and the acceptance threshold $\delta$ for accepting parameter values into the posterior distribution. We use $N_\text{pseu} = 10^5$ when inferring two parameters, and $N_\text{pseu} = 10^6$ when inferring six. When we use multiple objectives, we account for this by taking a weighted sum $d(\btheta)$ of our distance function $\tilde{d}(\btheta;E,T,C,D)$ in Eq.~(7) in the main manuscript and then accept $\btheta$ if this sum is below $\delta$, as below:
\begin{align}
    d(\btheta) = \frac{1}{N_\text{obj}}\sum_{i=1}^{N_\text{obj}} w_i \tilde{d}(\btheta;E_i,T_i,C_i,D_i) < \delta ~~&\Longrightarrow~~\text{accept $\btheta$ into the posterior.} \label{eq:d}
\end{align}
AABC and ABC studies \cite{thorneTopologicalApproximateBayesian2022,wenzel2025TopologicallybasedParameterInference,Lambert2018} occasionally define $\delta$ retroactively by enforcing that  $pN_\text{pseu}$ parameter values are accepted into the prior distribution. Here $0<p<1$ represents an acceptance fraction. We provide the conditions associated with our results by figure in the main manuscript below:
\begin{itemize}
\item \textbf{Figure 4:} The example patterns in panels (a), (b), and (d) are drawn from the data that we use to generate Fig.~6(a), Fig.~6(d), and Fig.~6(e), respectively. For wild type, we consider $(\alpha,\beta)$ given by $(2,4)$ for $d \approx 3$, $(4,2.5)$ for $d\approx 8$, $(2,2)$ for $d\approx 12$, and $(0,5)$ for $d \approx 26$. For \textit{pfeffer}, we consider $(\alpha,\beta)$ given by $(1.5,5.5)$ for $d\approx 3$, $(4.5,4)$ for $d\approx 5$, $(6.5,4)$ for $d\approx 8$, $(7,8.5)$ for $d\approx 15$, and $(3.5,0)$ for $d\approx 19$. For \textit{shady}, we consider $(\alpha,\beta)$ given by $(2.5,4)$ for $d\approx 1$, $(0.5,4.5)$ for $d\approx 3$, $(1.5,9.5)$ for $d\approx 6$, and $(2.5,2.5)$ for $d\approx 19$. Because \textit{nacre} patterns lack $M$ cells, altering $(\alpha,\beta)$ will have no effect. Instead, to show a wide range of distance values $d$, we generate example \textit{nacre} patterns by adjusting the parameters $(\gamma_1,\gamma_2,\tilde{a},\tilde{b},\tilde{c})$, where $\gamma_1$ and $\gamma_2$ are two parameters involved in $X^\text{d}$ and $X^\text{l}$ form transitions in the ABM \cite{volkeningIridophoresSourceRobustness2018}. In particular, they appear in the rules \cite{volkeningIridophoresSourceRobustness2018}:
\begin{align*}
\mathcal{N}(I^\text{l},\Omega^\textbf{x}_\text{loc}) > \gamma_1+ \mathcal{N}(I^\text{d},\Omega^\textbf{x}_\text{loc}) &\Longrightarrow \text{$X^\text{d}$ at $\textbf{x}$ swaps to $X^\text{l}$},\\
\mathcal{N}(I^\text{d},\Omega^\textbf{x}_\text{loc}) +P \mathcal{N}(X^\text{d},\Omega^\textbf{x}_\text{loc}) > \gamma_2 + \mathcal{N}(I^\text{l},\Omega^\textbf{x}_\text{loc}) + \mathcal{N}(M,\Omega^\textbf{x}_\text{loc}) &\Longrightarrow \text{$X^\text{l}$ at \textbf{x} swaps to $X^\text{d}$},
\end{align*}
where $P$ is a Bernoulli random variable with mean $0.5$. (Note that we simplify notation by using $\Omega_\text{loc}$ as shorthand for a few slightly different local neighborhoods, including balls with a radius of $45$ $\mu$m and balls with a radius of $75$ $\mu$m; see ref. \cite{volkeningIridophoresSourceRobustness2018} for full details on the length scales and rules. Broadly, these radii account for differences in cell--cell distances for various cell types \cite{volkeningIridophoresSourceRobustness2018}.)
In panel (b), we use $(\gamma_1,\gamma_2,\tilde{a},\tilde{b},\tilde{c}) = (2,1,3,9,4)$ for \textit{nacre} with $d\approx 2$; $(2,1,3,12,3)$ for $d\approx 7$; $(2,1,3,13,1)$ for $d\approx 18$; and $(4,4,1,1,1)$ for $d\approx 25$. Distances $d$ in Fig.~6 are rounded to the nearest integer; we note that panel (b) involves different random seeds (namely seeds $1,2,\dots,5$), and the other panels use our standard random seeds (seeds $2024001,2024002,\dots,2024005$). The ground-truth data have random seeds $10001,10002,\dots,10005$ in all cases.
    \item \textbf{Figure 6:} For two parameters involved in $M$ proliferation, Fig.\ 6 presents a series of examples, each considering only one objective ($N_\text{obj}=1$). Specifically, we set $w_1 = 1$ in \eqref{eq:d} above and use:
    \begin{itemize}[noitemsep,nolistsep]
    \item Fig.~6(a, f): $d(\btheta)=\tilde{d}(\btheta;\text{wild type}, M, 65,1)$; 
    \item Fig.~6(b, g): $d(\btheta)= \tilde{d}(\btheta;\text{wild type}, X^\text{d}, 46, 1)$; 
    \item Fig.~6(c, h): $d(\btheta)=\tilde{d}(\btheta;\text{wild type},M, 44, 1)$;
    \item Fig.~6(d, i): $d(\btheta)= \tilde{d}(\btheta;\textit{pfeffer}, M, 45, 1)$; 
    \item Fig.~6(e, j): $d(\btheta)=\tilde{d}(\btheta;\textit{shady},X^\text{d}, 65, 1)$.
    \end{itemize}
    We draw the heatmaps in Fig.~6(a)--(e) by evaluating these score functions at the $(\alpha,\beta)$-lattice points $(0.5x, 0.5y)$, where $x= 0,1, \dots, 20$ and $y = 0,1, \dots, 20$. Note that \textit{nacre} is not informative for inferring parameters in Fig.~6 because \textit{nacre} patterns lack $M$ cells. We use an acceptance threshold of $\delta = 1.2$ in all cases in Fig.~6. This corresponds to an acceptance percentage of $2.3\%$ for panel (f), $8.9\%$ for (g), $21.3\%$ for (h), $0.8\%$ for (i), and $0.3\%$ for (j). Applying a consistent acceptance threshold across these objectives allows us to highlight that different experimental conditions and timepoints lead to different ranges for $d(\btheta)$. This motivates our use of weights when we consider multiple objectives.
    \item \textbf{Figure 7:} This figure continues our study of two parameters involved in $M$ proliferation. In the top row of Fig.~7, we first repeat Fig.~6(f) and Fig.~6(j), and then overlay them in the rightmost panel of the top row as a visual guide. The second row of Fig.~7 considers multiple objectives, as below:
    \begin{itemize}[noitemsep,nolistsep]
    \item Fig.~7 (bottom left): We use $N_\text{obj}=4$, with $(E_i,T_i,C_i,D_i)$ given by $ (\text{wild type},M,65,1)$,\\ $(\text{wild type},X^\text{d}, 46, 1)$, $(\textit{pfeffer},M, 45, 1)$, and $(\textit{shady},X^\text{d}, 65, 1)$, for $i=1,2,3,4$, respectively. These objectives correspond to two final simulation times and two times roughly in the middle of the developmental timeline that we consider. As in Fig.~6, we exclude \textit{nacre} experiments (i.e., patterns lacking $M$ cells) because these are uninformative for $M$ parameters. This example allows us to illustrate how a small set of carefully chosen objectives can lead to practical identifiability. In addition to final times (65 dpf), the middle times are valuable because they correspond to the point at which the first two wild-type stripes or first two rows of \textit{pfeffer} spots are complete and patterning has progressed to the next stage (e.g., initiating gold interstripes in wild type).
    \item Fig.~7 (bottom right): We use a large set of $N_\text{obj}=80$ objectives, encompassing $(E_i,T_i,C_i,D_i)$ given by $(\text{wild type}, M, t,1)$ for $t=44,\dots,65$, $(\text{wild type}, X^\text{d},t,1)$ for $t=44,\dots,65$, $(\textit{pfeffer}, M,t,1)$ for $t=40,\dots,65$, and $(\textit{shady}, X^\text{d},t,1)$ for $t=56,\dots,65$. These objectives represent the second half of the patterning timeline, which we suggest is more informative because cell behaviors are less driven by the initial condition. We note that \textit{shady} patterns form more slowly than other patterns in the ABM \cite{volkeningIridophoresSourceRobustness2018}, so we consider a narrower time range there.
    \end{itemize}
   We accept parameter values according to \eqref{eq:d} with weights $\{w_i\}$ defined in Eq.~(9) in the main manuscript, and we choose $\delta$ to achieve a $1\%$ acceptance percentage. This corresponds to $\delta = 3.44$ in \eqref{eq:d} for the four-objectives case and $\delta = 3.14$ for the 80-objectives case. Together Fig. 6 and Fig. 7 illustrate two methods for enforcing a consistent acceptance parameter across different examples: we can set the threshold $\delta$ to be constant, or we can set the acceptance percentage $p$ to be constant.
    \item \textbf{Figure 8(a)--(d):} This figure contains our case study of $I^\text{d}$ and $I^\text{l}$ movement, making \textit{shady} patterns, which lack iridophores, uninformative. For panels (a)--(d), we consider a few example single objectives ($N_\text{obj}=1$) as below:
    \begin{itemize}[noitemsep,nolistsep]
    \item Fig.~8(a): $d(\btheta) = \tilde{d}(\btheta;\text{wild type},X^\text{d},43,1)$;
    \item Fig.~8(b): $d(\btheta) = \tilde{d}(\btheta;\text{wild type},M,65,1)$;
    \item Fig.~8(c): $d(\btheta) = \tilde{d}(\btheta;\text{wild type},X^\text{d},65,1)$;
    \item Fig.~8(d): $d(\btheta) = \tilde{d}(\btheta;\textit{pfeffer},M,65,1)$.
    \end{itemize}
    To illustrate another means of enforcing a consistent acceptance threshold across examples, here we use $\delta = 7$ and weight $w_1$ in \eqref{eq:d} specified as the minimum value of each score function over the pre-proposals (see Eq.~(9) in the main manuscript); this means panels (a)--(d) in Fig.~8 each involve different weights $w_1$. This corresponds to an acceptance percentage of $17.9\%$ for panel (a), 
    $16.4\%$ for (b), $19.8\%$ for (c), and $54.5\%$ for (d).
    \item \textbf{Figure 8(e):} Continuing our study of $I^\text{d}$ and $I^\text{l}$ migration parameters, we consider a multi-objective example in Fig.~8(e). Specifically, we use $N_\text{obj} = 29$ objectives $(E_i,T_i,C_i,D_i)$ with $(\text{wild type},M,t,0)$ for $t =45,58,65$;
    $(\text{wild type},M,t,1)$ for $t =42,62,65$;
    $(\text{wild type},X^\text{d},t,0)$ for $t =41,42,45,65$;
    $(\text{wild type},X^\text{d},t,1)$ for $t =43,47,65$;
    $(\text{wild type},X^\text{l},65,0)$;
    $(\text{wild type},X^\text{l},t,1)$ for $t =45,54,65$;
    $(\textit{pfeffer},M,t,0)$ for $t =56,65$;
    $(\textit{pfeffer},M,65,1)$;
    $(\textit{nacre},X^\text{d},t,0)$ for $t =40,44,65$;
    $(\textit{nacre},X^\text{d},t,1)$ for $t =44,65$;
    $(\textit{nacre},X^\text{l},t,0)$ for $t =42,65$;
    $(\textit{nacre},X^\text{l},t,1)$ for $t =45,65$. We select these objectives through visual inspection of the associated heatmaps. To be consistent with the single-objective results in Fig.~8(a)--(d), we apply \eqref{eq:d} with an acceptance threshold of $\delta = 7$, leading to an acceptance percentage of $0.13\%$.
    \item \textbf{Figure 9:} Because our last case study involves rule inference across $81$ candidate models and inspecting the posteriors for six-dimensional $\btheta =(a,b,c,\tilde{a},\tilde{b},\tilde{c})$ under many alternative objectives is intractable, we use a non-selective combination of $N_\text{obj}=192$ objectives in Fig.~9. These objectives $(E_i,T_i,C_i,D_i)$ are $(\text{wild type},M,t,\text{dim})$ for $t=44,\dots,65$ and $\text{dim} = 0,1$; $(\text{wild type},X^\text{d},t,\text{dim})$ for $t=44,\dots,65$ and $\text{dim} = 0,1$; $(\textit{pfeffer},M,t,\text{dim})$ for $t=40,\dots,65$ and $\text{dim} = 0,1$; and $(\textit{nacre},X^\text{d},t,\text{dim})$ for $t=40,\dots,65$ and $\text{dim} = 0,1$. This corresponds to all of our objectives for $M$ and $X^\text{d}$ cells, excluding earlier timepoints that tend to be less informative, and \textit{shady} patterns because they lack iridophores. For example, $I^\text{l}$-to-$I^\text{d}$ transitions do not occur in wild type until around $44$~dpf \cite{volkeningIridophoresSourceRobustness2018}. (We also exclude persistence landscapes generated using $X^\text{l}$ cells for simplicity.)
\end{itemize}

\begin{figure}
\centering
\includegraphics[width=\textwidth]{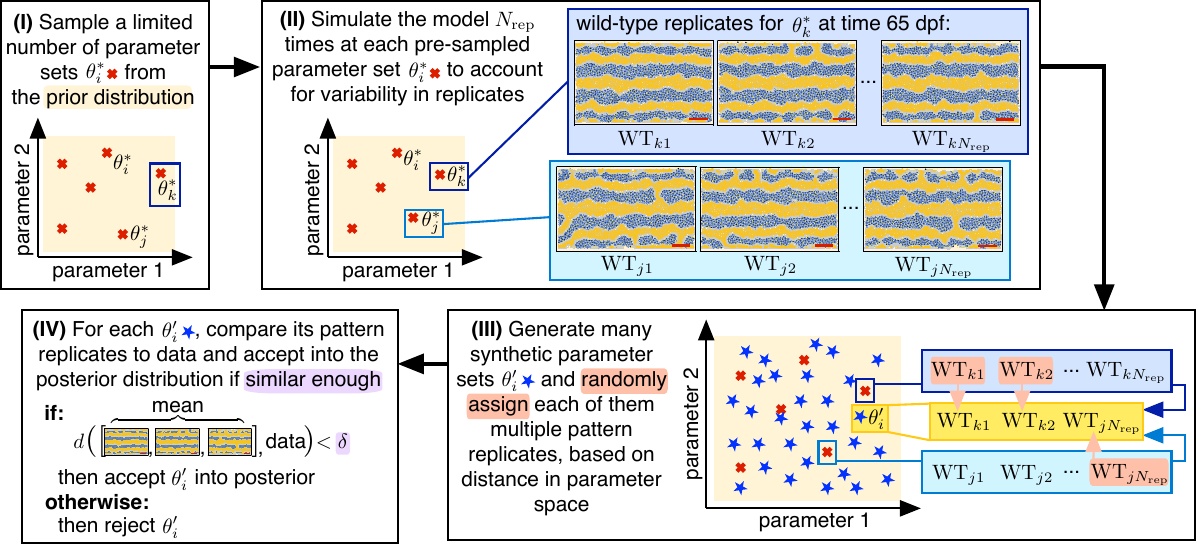}\vspace{-\baselineskip}
\caption{Single-objective AABC in the context of zebrafish patterning. Developed by Buzbas and Rosenberg \cite{buzbasAABCApproximateApproximate2015}, AABC is a means of estimating the parameter posterior distribution for computationally expensive, stochastic models \cite{Lambert2018} The pre-proposal stage of AABC, Steps~(I)--(II), involves sampling a limited number $N_\text{pre}$ of parameter values from the prior distribution and simulating our model multiple times per parameter value $\btheta^*$. In this example, the objective is to replicate wild-type patterns at the final simulation time ($65$~dpf). In the pseudo-proposal stage of AABC, Step~(III), we again sample from the prior, this time generating $N_\text{pseu} \gg N_\text{pre}$ possible parameter sets $\btheta'$. Instead of simulating the ABM \cite{volkeningIridophoresSourceRobustness2018} again at each $\btheta'$, we reuse the data generated in Step~(2), randomly assigning each $\btheta'$ some pattern replicates drawn from its $N_\text{nei}$ neighboring pre-proposals $\btheta^*$ (we use $N_\text{nei} = N_\text{rep}$). AABC concludes with the acceptance stage, Step~(IV), when we quantitatively compare the patterns associated with $\btheta'$ with our data to determine if we will accept $\btheta'$ into the posterior distribution. Choosing a metric $d(\cdot)$ to compare variable, messy spatial patterns to data is challenging in general, and our work addresses this issue through TDA and persistence landscapes \cite{bubenik2015StatisticalTopologicalData,bubenik2017PersistenceLandscapesToolbox,Hartsock2025}.}
    \label{fig:aabc_schematic}
\end{figure}

\begin{figure}
    \centering
\includegraphics[width=\textwidth]{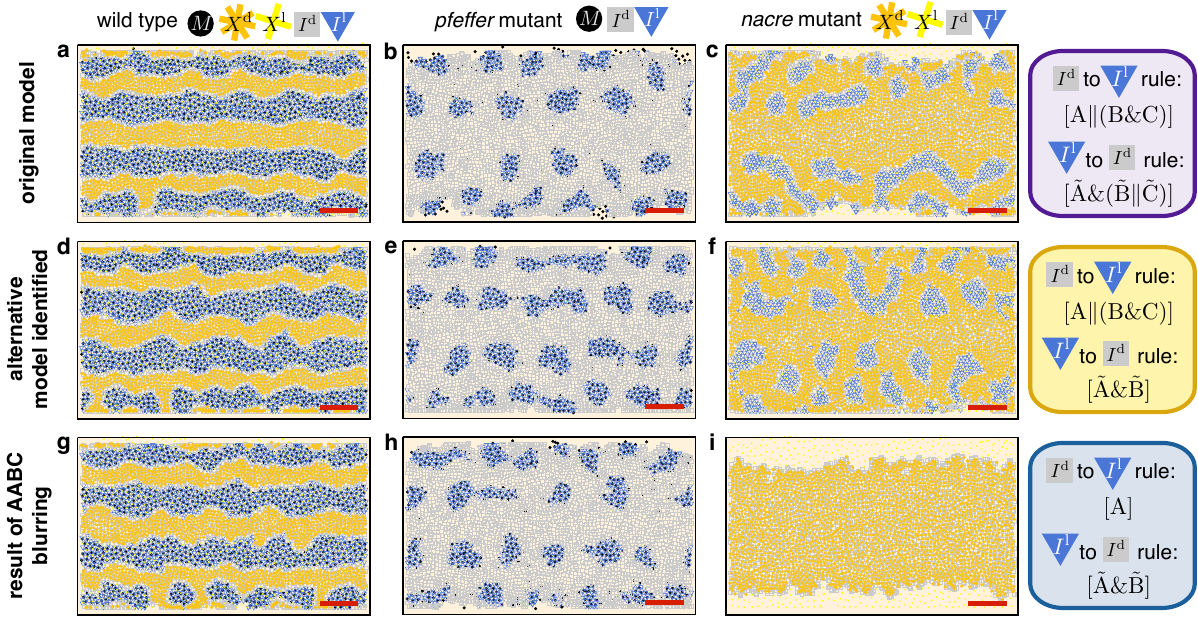}\vspace{-\baselineskip}
\caption{Example patterns under alternative models. We show (a) wild-type, (b) \textit{pfeffer}, and (c) \textit{nacre} patterns under the original cell-type transition rules for $I^\text{d}$ and $I^\text{l}$ in the ABM \cite{volkeningIridophoresSourceRobustness2018}. (d)--(f) Our multi-objective inference methodology identifies a simpler model (associated with high $\tilde{c}$), in which the original rule for $I^\text{l}$-to-$I^\text{d}$ transitions is replaced by $[\tilde{\text{A}} \& \tilde{\text{B}}]$. We show example simulations for the resulting (d) wild-type, (e) \textit{pfeffer}, and (f) \textit{nacre} conditions. While the \textit{nacre} pattern in panel (f) has some differences from the one in panel (c), both feature an expanded central orange region with rough boundaries and blue regions, and we note that \textit{nacre} patterns are highly variable \cite{volkeningIridophoresSourceRobustness2018,mcguirlTopologicalDataAnalysis2020}. Importantly, Volkening and Sandstede \cite{volkeningIridophoresSourceRobustness2018} highlighted that iridophores must rely on signals associated with $M$ and $X^\text{d}$, as iridophores are present in both the $I^\text{d}$ and $I^\text{l}$ forms in \textit{pfeffer} (lacking $X^\text{d}$ and $X^\text{l}$) and \textit{nacre} (lacking $M$) patterns. The alternative model that we identify in panels (d)--(f) agrees with this observation \cite{volkeningIridophoresSourceRobustness2018}. (g)--(i) On the other hand, our pipeline can also generate false alternative models due to the blurring process in AABC \cite{buzbasAABCApproximateApproximate2015} that helps reduce computational cost. Here we show example simulations from a false alternative model that we identify using our pipeline. While the resulting (g) wild-type and (h) \textit{pfeffer} patterns look qualitatively similar to our target patterns, (i) \textit{nacre} does not. This matches the intuition in \cite{volkeningIridophoresSourceRobustness2018}: reducing the $I^\text{d}$-to-$I^\text{l}$ rule to simply $[\text{A}]$ does not work because $I^\text{d}$ cells then rely on signals from $M$ only to change forms, meaning that that no $I^\text{l}$ cells appear in \textit{nacre}. All patterns are shown at $65$ dpf; panels (a)--(c) use the baseline parameters from \cite{volkeningIridophoresSourceRobustness2018}, namely $(a,b,c,\tilde{a},\tilde{b},\tilde{c}) = (3,5,2,3,9,3)$ cells; panels (d)--(f) use $(a,b,c,\tilde{a},\tilde{b},\tilde{c}) = (2,2,2,3,11,13)$ cells; and panels (g)--(i) use $(a,b,c,\tilde{a},\tilde{b},\tilde{c}) = (3,-1,0,3,11,10)$ cells. (We note that $(a,b,c,\tilde{a},\tilde{b},\tilde{c}) = (3,-1,0,3,11,3)$ cells, which corresponds to $[\text{A}]$ and the original rule $[\tilde{\text{A}} \& (\tilde{\text{B}} \| \tilde{\text{C}})]$, produces similarly poor \textit{nacre} patterns, suggesting that the issue in panel (i) is due to $[\text{A}]$.) }
\end{figure}

\begin{figure}
    \centering
\includegraphics[width=\textwidth]{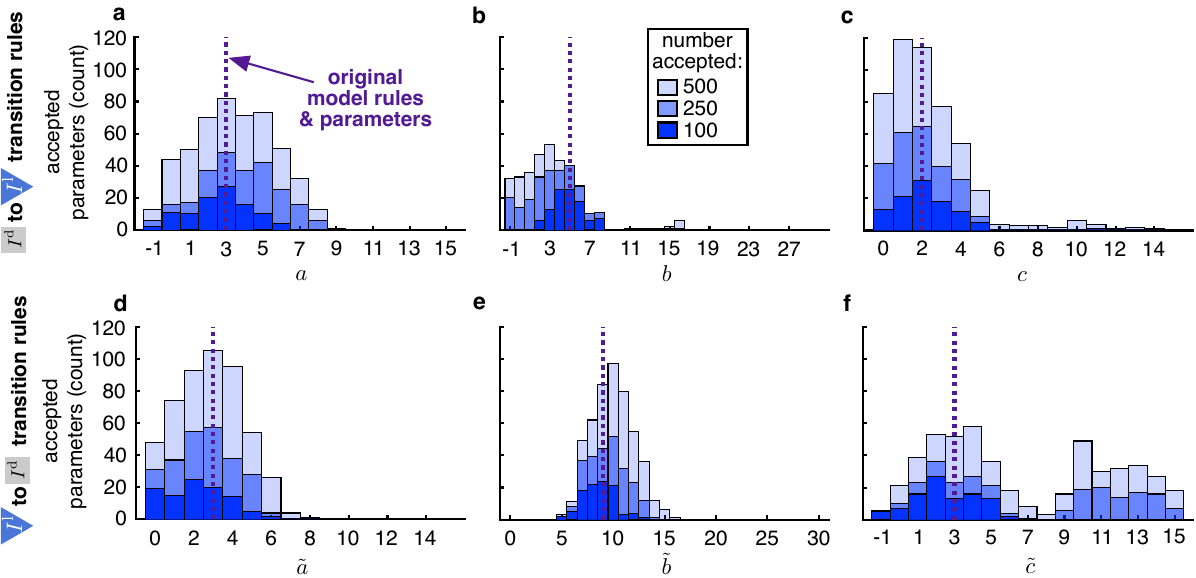}\vspace{-\baselineskip}
\caption{Effects of the acceptance threshold $\delta$ in AABC. Also see Fig.~9 in the main manuscript. Marginal posterior distributions for (a) $a$, (b) $b$, (c) $c$, (d) $\tilde{a}$, (e) $\tilde{b}$, and (f) $\tilde{c}$ highlight that changing the threshold for accepting parameter values during AABC \cite{buzbasAABCApproximateApproximate2015} changes qualitative features of the resulting posterior distribution. Most notably, we lose the double peak in panel (f) that alerts us to the alternative model in Fig.~S2(d)--(f). The acceptance threshold $\delta$ is often selected in a largely \textit{ad hoc} way in modeling studies that use AABC or ABC, and this process can involve tedious and subjective manual evaluation, especially when considering multiple objectives. To gain further insight, we sample $500$ parameter values from the distributions in panels (a)--(f), and manually evaluate their performance on wild-type, \textit{nacre}, and \textit{pfeffer} patterns at $65$ dpf, classifying them as ``good" if they perform well on all three patterns. (We note that this manual process is imperfect and subjective, and it is limited to final times and single stochastic replicates.) According to this process, we find that about $7\%$ of the parameters are good when we draw from the distribution that we generate by selecting $\delta$ so that $500$ out of $10^6$ parameter values are accepted (light blue distributions); about $13\%$ are good when we accept only $250$ parameter values (medium blue distributions); and about $20$\% are good when we accept only $100$ parameters (dark blue distributions). Moreover, reducing the acceptance criteria drastically to accept only $20$ parameter values out of $10^6$ leads to $25\%$ good parameters based on our manual evaluation. We suggest that this illustrates the effects of blurring in AABC and highlights the tradeoff between computational cost and posterior resolution, particularly when considering our very broad prior distribution that spans $81$ candidate models. In our case, because the posterior predictive check shows that some parameter values in the second peak of the $\tilde{c}$ distribution in panel (f) lead to good patterns, we specify $\delta$ so that $250$ parameter values are accepted in the main manuscript. Our work provides a first step toward rule inference in detailed ABMs, and, as we highlight in the Discussion, we suggest that developing data-driven methods to inform the choice of the acceptance threshold $\delta$ in AABC (as well as other hyperparameters such as $N_\text{rep}$ and $N_\text{nei}$; see Methods) is an important direction for future work.}
\end{figure}

\end{document}